\documentclass[]{emulateapj}

\newcommand\bb[1]{\mbox{\boldmath{$#1$}}}
\usepackage{hyperref}
\usepackage{xcolor}
\usepackage{amsmath}
\definecolor{dark-red}{rgb}{0.4,0.15,0.15}
\definecolor{dark-blue}{rgb}{0.15,0.15,0.4}
\definecolor{medium-blue}{rgb}{0,0,0.5}
\hypersetup{
    colorlinks, 
    linkcolor={blue},
    citecolor={blue}, 
    urlcolor={blue}
}
\usepackage[varg]{txfonts}

\slugcomment{Draft version \today}

\shorttitle{\textsc{Anisotropic Nature of MRI-driven Turbulence}}
\shortauthors{\textsc{Murphy \& Pessah}}

\usepackage{color}
\definecolor{brown}{rgb}{0.42,0.24,0.07}
\definecolor{darkgreen}{rgb}{0.0,0.6,0.00}
\definecolor{purple}{rgb}{0.7,0.0,0.7}

\begin{document}

\title{On The Anisotropic Nature of MRI-driven Turbulence in Astrophysical Disks}

\author{Gareth C. Murphy and Martin E. Pessah}
\affil{Niels Bohr International Academy, Niels Bohr Institute, Blegdamsvej 17, 2100 Copenhagen, Denmark}
                 \email{gmurphy@nbi.dk, mpessah@nbi.dk}

\begin{abstract}
The magnetorotational instability is thought to play an important role
in enabling accretion in sufficiently ionized astrophysical disks.
The rate at which MRI-driven turbulence transports angular momentum 
is intimately related to both the strength of the amplitudes
of the fluctuations on various scales and the degree of anisotropy of
the underlying turbulence.  This has motivated several studies to
characterize the distribution of turbulent power in spectral space.  In this
paper, we investigate the anisotropic nature of MRI-driven turbulence
using a pseudo-spectral code and introduce novel ways for providing a
robust characterization of the underlying turbulence.  We show that the
general flow properties vary in a quasi-periodic way on timescales
comparable to $\sim\!\!10$ inverse angular frequencies motivating the
temporal analysis of its anisotropy.  We introduce a 3D tensor
invariant analysis to quantify and classify the evolution of the
anisotropy of the turbulent flow. This analysis shows a
continuous high level of anisotropy, with brief sporadic transitions
towards two-component and three-component isotropic turbulent flow.
This temporal-dependent anisotropy renders standard shell-average,
especially when used simultaneously with long temporal averages,
inadequate for characterizing MRI-driven turbulence.  We propose an
alternative way to extract spectral information from the turbulent
magnetized flow, whose anisotropic character depends strongly on time.
This consists of stacking 1D Fourier spectra along three orthogonal
directions that exhibit maximum anisotropy in Fourier space.  The
resulting averaged spectra show that the power along each of the three
independent directions differs by several orders of magnitude over
most scales, except the largest ones.  Our results suggest that a
first-principles theory to describe fully developed  MRI-driven 
turbulence will likely have to consider the anisotropic nature of the flow 
at a fundamental level.
\end{abstract}


\section{Introduction}

The molecular viscosity operating in a laminar flow is too small by many
orders of magnitude in order to account for the rate of angular
momentum transfer inferred from observations of astrophysical disks
\citep{2002apa..book.....F}. It has long been thought that turbulence
offers a mechanism for efficient angular momentum transport in these
disks \citep{Shakura:1973uy,LyndenBell:1974uq}. 
Regardless of its nature,
turbulence in accretion disks must be anisotropic, since strictly isotropic 
turbulence is unable to lead to angular momentum transport.

In sufficiently ionized disks, the magnetorotational instability
\citep[MRI; ][]{1991ApJ...376..214B} offers a mechanism to drive
magnetohydrodynamic turbulence
\citep{Hawley:1995gd,Brandenburg:1996du}.  In its most simple and
powerful form, the MRI is a linear instability mediated by a weak
vertical magnetic field that taps into the free energy of the
differentially rotating flow leading to exponential growth of the
radial and azimuthal components of the velocity and magnetic fields.
The correlated growth of these field components leads to transport of
angular momentum mediated by Reynolds and Maxwell stresses, with the
latter being the dominant, by a factor of a few  in ideal MHD, both in the linear
phase of the instability and the ensuing turbulence 
\citep{2006MNRAS.372..183P}.

There are two important properties of MRI-driven turbulence which influence
 the rate at which angular momentum is transported. One
is clearly the strength of the fluctuations in the velocity and
magnetic fields.  Equally important, however, is the degree of
anisotropy exhibited by the stress tensors associated with them, since completely 
isotropic tensors would lead to vanishing angular momentum transport.  
In the ideal MHD limit, the radial and azimuthal components of both the 
velocity and magnetic fields are exactly equal in the linear phase of 
the MRI and dominate over the vertical component. In the turbulent state,
the azimuthal components dominate over their radial and vertical
counterparts. This naturally creates a strong anisotropy in the
turbulent flow. This suggests that the extent to which the flow is
anisotropic evolves significantly from the linear phase of the MRI to
the nonlinear, turbulent regime.  Although there have been many studies of the
linear phase of the MRI and its non-linear evolution, only a
relatively small fraction have explored in detail the mechanism
responsible for its saturation and none have focused explicitly on the
evolution of the degree of anisotropy exhibited by the magnetized flow as it 
evolves from the linear regime of the instability to the ensuing turbulent state.

Motivated by the need to understand the processes at work, in this
paper we analyze the saturation of the MRI and the ensuing anisotropic
turbulence by means of targeted numerical experiments.
This enables us to follow closely the evolution of the magnetized fluid
through three main stages: the linear regime where the MRI grows
unimpeded, the early non-linear evolution leading to the turbulent
state, and the fully developed turbulent regime.  Aided by these
simulations, we investigate how the amplitude of the fluctuations and
the degree of anisotropy evolve throughout by employing statistical
tools that have not been applied to MRI-driven turbulence before.

In order to extract statistical information about the properties of
MRI-driven turbulence, most previous works have focused on using spherical 
shell averages in spectral space, which is suitable for isotropic turbulence. 
However, the underlying spectral distributions in MRI-driven turbulence are
far from isotropic. This suggests that the use of spherical shell averages may
not be optimal for analyzing and representing spectral energy
distributions in MRI-driven turbulence. As a viable alternative, we employ a joint approach
which consists of using invariant maps for measuring anisotropy in
real space and dissecting the three-dimensional Fourier spectrum along the 
most relevant planes, as revealed by the anisotropic nature of the flow. 
We envisage that the new insights offered by this approach will be valuable 
for guiding a more fundamental understanding of the anisotropic nature of
magnetohydrodynamic turbulence in astrophysical disks.

The paper is organized as follows. In section \ref{previous}, we
provide an overview of previous works devoted to analyzing the
mechanisms that lead to the saturation of the MRI. We motivate 
the use of targeted numerical experiments to shed light into the
processes involved when the agent responsible for saturation is
related to parasitic instabilities. In section \ref{exact}, we examine
the KH instability of a sinusoidal velocity field in order to prepare
the ground to study the development and saturation of the MRI due to
secondary instabilities in Section \ref{latelinearMRI}.  In Section \ref{anisotropy},
we introduce the concepts involved in invariant analysis introduced by
\cite{1977JFM....82..161L} and use them to study the evolution of anisotropy 
of the MRI-driven turbulent flow.  As a complementary approach, we analyze the 
evolution of the three-dimensional spectral distribution characterizing the energy density
and the Maxwell stress in Fourier space, from the early phases of the non-linear 
stage of the MRI to the fully developed turbulent state. 
By carefully stacking power spectra along the principal directions characterizing the
turbulent state, we define a more meaningful way of describing its anisotropic nature. 
We summarize our results and discuss their implications in Section  \ref{discussion}.

\section{Previous Works on the Saturation of the MRI}
\label{previous}

Several different approaches have been taken in the literature to
investigate the mechanisms responsible for the saturation of the MRI.
One group of works have focused on understanding the role of parasitic
instabilities that feed off the primary MRI mode in the incompressible
MHD regime using analytical
\citep{Goodman:1994dd,Pessah:2009gm,Pessah:2010ic}; and numerical
\citep{Latter:2009br,2010A&A...516A..51L} means; while others
\citep{Obergaulinger:2009fv,Latter:2009br} have relaxed the assumption
of an incompressible flow and used numerical simulations to explore
how the instability saturates.  All of these studies have been carried
out in the framework of the shearing-sheet where, by definition, the
background flow, and in particular the shear profile feeding the MRI,
remains unaltered. This is probably a reasonable assumption in
astrophysical settings where the ultimate source of energy responsible
for differential rotation (i.e., the central potential) remains unchanged.

A second approach has focused on the saturation of the MRI due to
flattening of the background shear profile. This route to suppress the
energy source for the instability is present in settings where the
shear profile is not imposed throughout the flow, as it is the case
when rotating rigid walls are considered. A series of papers by
\citet{2005PhFl...17i4106K,2006EAS....21...81J,Jamroz:2008fn,
Jamroz:2008jw,Tatsuno:2008fv,Julien:2010cn} find that in laboratory
settings, the MRI saturated state is not turbulent, but consists of
laminar axisymmetric layers threaded by magnetic fields with
alternating polarity (see, e.g., Figure  4 in
\citealt{Jamroz:2008jw}). The initial background shear profile is
modified and approaches solid body rotation leading to a laminar
saturated state that settles on a resistive timescale.
\citet{Tatsuno:2008fv} carried out two-dimensional simulations with
rigid boundaries, finding that the initial Keplerian background
profile becomes flattened (see, e.g., their Figure  5). In this study,
parasite modes related to Kelvin Helmholtz instabilities are not
detected, although they do detect tearing modes that lead to the
formation of magnetic islands between laminar magnetic layers.

Other approaches have focused on saturation via diversion of energy
into magnetosonic waves, or generation of large-scale
magnetic fields via a dynamo process.  \citet{2012PhRvL.109v4501L}
suggested that the MRI saturates by driving non-resonant magnetosonic
waves in a bursty oscillatory manner.  \citet{Umurhan:2007hs} perform
a weakly non-linear analysis of the MRI and find that it saturates with only
a modest reduction in background shear, with an amplitude proportional
to the square root of the magnetic Prandtl number, provided this is much
smaller than unity. \citet{Vishniac:2009il} suggested that, in the
absence of net magnetic flux, the MRI dynamo saturation level is
controlled by the imposition of a maximum wavenumber on MRI modes due
to the combined effects of magnetic tension, buoyancy and/or
resistivity or viscosity. Alternatively, \citet{Ebrahimi:2009ey} 
proposes that the MRI saturates by generating large scale fields 
through an $\alpha-$effect, although they point out that, since the total toroidal 
and axial flux does not change significantly with time, this cannot be referred 
to as a large-scale dynamo effect. 

In this work, we focus on parasite modes as a potential cause of saturation.  
The notion that the exponentially growing MRI modes are themselves in turn 
subject to secondary, parasitic instabilities was originally suggested by \citet{Goodman:1994dd} 
and has been considered in more detail by a number of works more recently
\citep{2008A&A...487....1B,Obergaulinger:2009fv,Pessah:2009gm,Latter:2009br,Pessah:2010ic}.
The physical arguments that support the development of secondary instabilities
are quite robust. Provided that the initial magnetic field is very weak, 
the MRI modes are exact solutions of the equations of motion and thus they can 
reach large amplitudes. These modes lead to exponentially growing velocity and
magnetic fields which vary sinusoidally in the vertical direction (perpendicular to the
disk midplane). When the amplitude of these perturbations is large enough, they 
become themselves subject to secondary instabilities. 
Broadly speaking, these secondary parasitic modes belong to two classes
related to the Kelvin-Helmholtz (KH) instability and the tearing mode
instability (TM). As discussed in \citet{Pessah:2010ic}, the KH
parasitic mode dominates in regimes near ideal MHD, appropriate for
highly ionized accretion disks, whereas the TM parasite dominates in
highly resistive regimes, such as laboratory experiments and
protoplanetary disks.  Since these secondary instabilities extract
energy from the MRI, they are expected to limit its growth and to
potentially cause its saturation.

Because of the complexity inherently involved in carrying out secondary
instability analyses, previous semi-analytical works have made a
number of approximations.  In particular, the exact (primary) MRI
modes are considered as a time-independent background from which the
(secondary) parasitic modes feed off. The effects of the weak vertical
background field, the Coriolis force, and the background shear flow,
are all usually neglected in the equations of motion governing the
dynamics of the secondary instabilities \citep{Goodman:1994dd,Pessah:2009gm}.  
Relaxing these approximations
in analytical studies is a challenging task, therefore we have
designed tailored numerical experiments which allow us to follow in
detail the evolution of the flow from the early onset of the
instability until the fully development turbulent regime ensues.
The numerical experiment that we consider below isolates the fastest growing, 
axisymmetric MRI mode and follows its evolution after seeding the perturbations
that excite a non-axisymmetric KH parasitic mode of the type that 
are expected to exhibit the fastest growth according to semi-analytical studies
\citet{Pessah:2010ic}. This controlled experiment allows to us address
the evolution of the degree of anisotropy in the flow from the early
linear phase of the MRI until the fully developed turbulent state.

\begin{figure}[t]
\begin{center}
\includegraphics[width=0.5\textwidth]{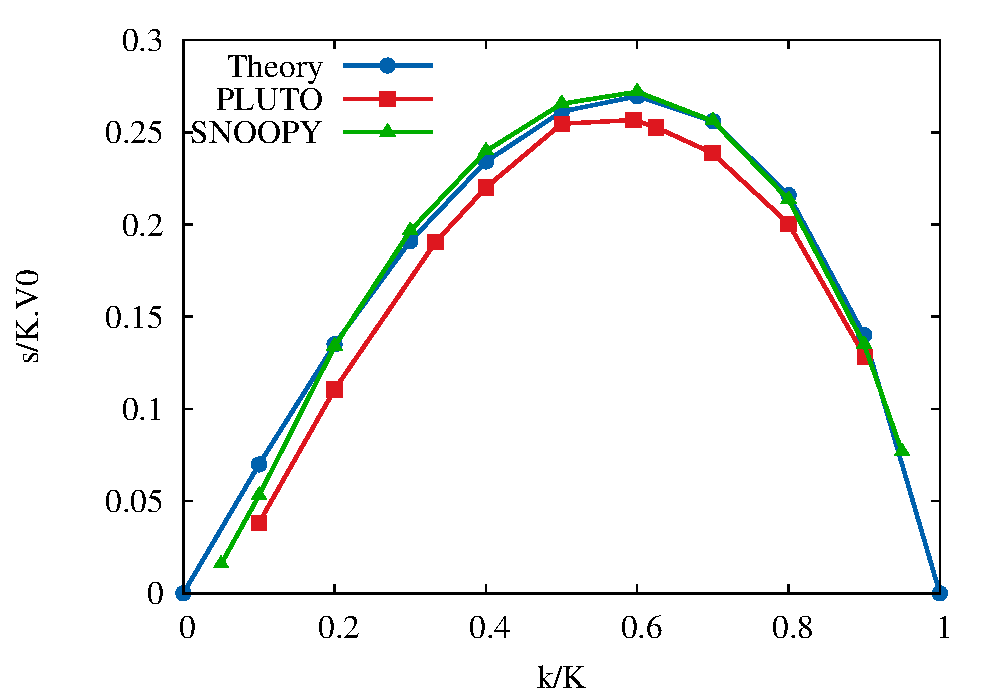}
\caption{
Comparison of the dimensionless growth rates for KH instabilities
excited by a sinusoidal velocity profile as a function of $k_h/K$,
i.e., the ratio of the wavenumbers associated with the periodic background
shear flow and the perturbations.  The pseudo-spectral code, SNOOPY agrees
most closely with the theoretical results derived for an
incompressible flow.  The results obtained with the PLUTO code show
that the effect of compressibility is to slightly reduce the growth
rates expected from analytical theory.}
\label{kh2danalytical_simcomparison}
\end{center}
\end{figure}

\section{Exact MRI Modes and Related KH Parasites}
\label{exact}

\subsection{Equations and fastest growing modes}

The equations governing the local dynamics of a magnetized disk
in the shearing box framework  are given by
\citep{Hawley:1995gd,Brandenburg:1996du}
\begin{align}
&\frac{\partial  \bb{v}}{\partial t}
+\bb{v} \cdot \nabla \bb{v} 
=
- 2 \bb{\Omega} \times \bb{v} \, + 
\, 2 q \Omega^2 x 
- \frac{\nabla p}{\rho}
+ \frac{\bb{J}\times \bb{B}}{\rho} + \nu \nabla^2 v \,,
\label{eq:momentum}
\\
&\frac{\partial \mathbf{B} }{\partial t}
= \nabla
\times
\left(
\bb{v} \times \bb{B}
\right)
+
\eta \nabla^2 \bb{B} \,.
\label{eq:induction}
\end{align}
Here, $\rho$ is the mass density, $\bb{v}$ is the fluid velocity in
the rotating frame, with angular frequency $\bb{\Omega} = \Omega
\,\hat{\bb{z}}$, $p$ is the pressure determined through an equation of
state, and $\bb{B}$ is the magnetic field, with $\nabla \cdot {\bb
B}=0$.  The current density is $\bb{J}\equiv \nabla\times \bb{B}/
\mu_0$, with $\mu_0$ a constant dependent on the unit system adopted.
The first and second terms on the right hand side of equation
(\ref{eq:momentum}) account for the Coriolis acceleration and the
first order expansion of the tidal potential. The shear parameter is
defined as
\begin{equation}
q= - \frac{d \ln{\Omega}}{d \ln{r}} \,.
\end{equation}
so that $q=3/2$ for a Keplerian disk. The coefficients $\nu$ and $\eta$
stand for the viscosity and resistivity, assumed to be constant.
    
Let us consider an incompressible flow, which is a good
approximation provided that the magnetic field involved is very
weak, i.e., the magnetic energy density is small compared to its
thermal counterpart.  The set of equations
(\ref{eq:momentum})--(\ref{eq:induction}) admits exact, MRI-unstable
solutions of the form
\begin{eqnarray}
\bb{v}&=&- q \Omega x \check{\bb{y}} +  \bb{V}_0 \sin(Kz) \, e^{\Gamma t} \,,\\ 
\bb{B}&=& \bar{B}_z \check{\bb{z}} +  \bb{B}_0 \cos(Kz) \, e^{\Gamma t} \,.
\end{eqnarray}
The ratio of the amplitudes, i.e., $B_0/V_0$, sets the initial amplitude of the MRI mode.
In ideal MHD, the wavelength of the fastest mode is
$\lambda_{max} = 2\pi K_{\rm max} = \sqrt{15/16} \, \bar{v}_{{\rm A},
z}/\Omega$, where $\bar{v}_{{\rm A}, z}=\bar{B}_z/\sqrt{4\pi \rho}$ is
the Alfv\'en speed associated with the background magnetic field, $\bar{B}_z$.
This mode grows at the rate $\Gamma_{\rm max} = 3\Omega/4$, with
$V_{A,0}/V_0 = \sqrt{5/3}$, where $V_{A,0}=B_0/\sqrt{4\pi \rho}$ is
the Alfv\'{e}n speed associated with the MRI-driven perturbation
\citep{2006MNRAS.372..183P}.

These exact primary MRI modes are themselves in turn subject to
secondary, parasitic instabilities. The fastest growing parasitic
modes, which are associated with KH modes, have the same vertical
periodicity as the primary MRI mode itself
\citep{Goodman:1994dd,Pessah:2009gm}.  As a prelude to analyzing in
detail the non-linear evolution of the MRI modes as they are affected
by secondary instabilities, we perform a series of simpler numerical
experiments to study the dynamics of the KH instabilities resulting
from a periodic velocity profile with constant amplitude.

\begin{figure}[t]
\begin{center}
\includegraphics[width= 0.49\columnwidth]{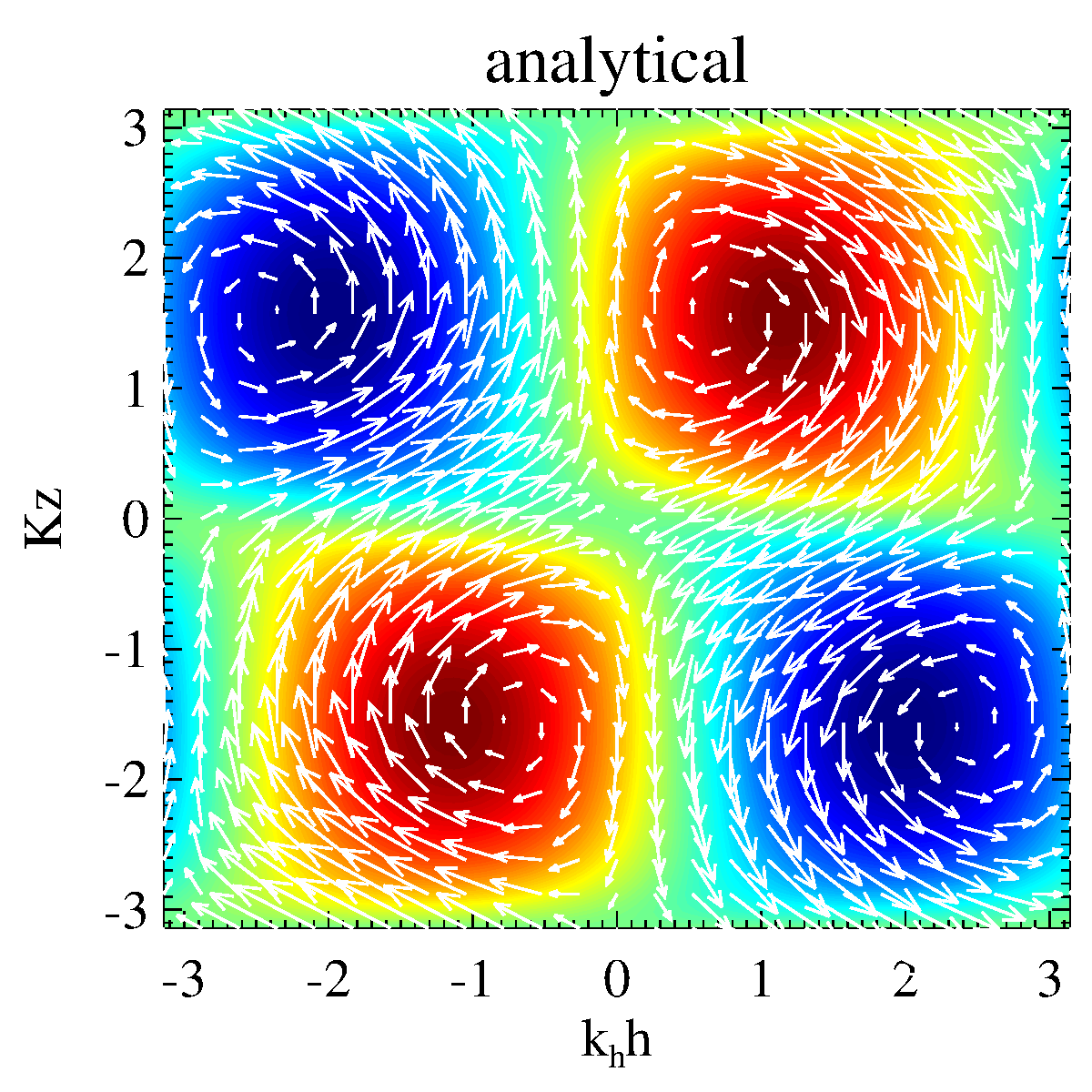}  
\includegraphics[width= 0.49\columnwidth]{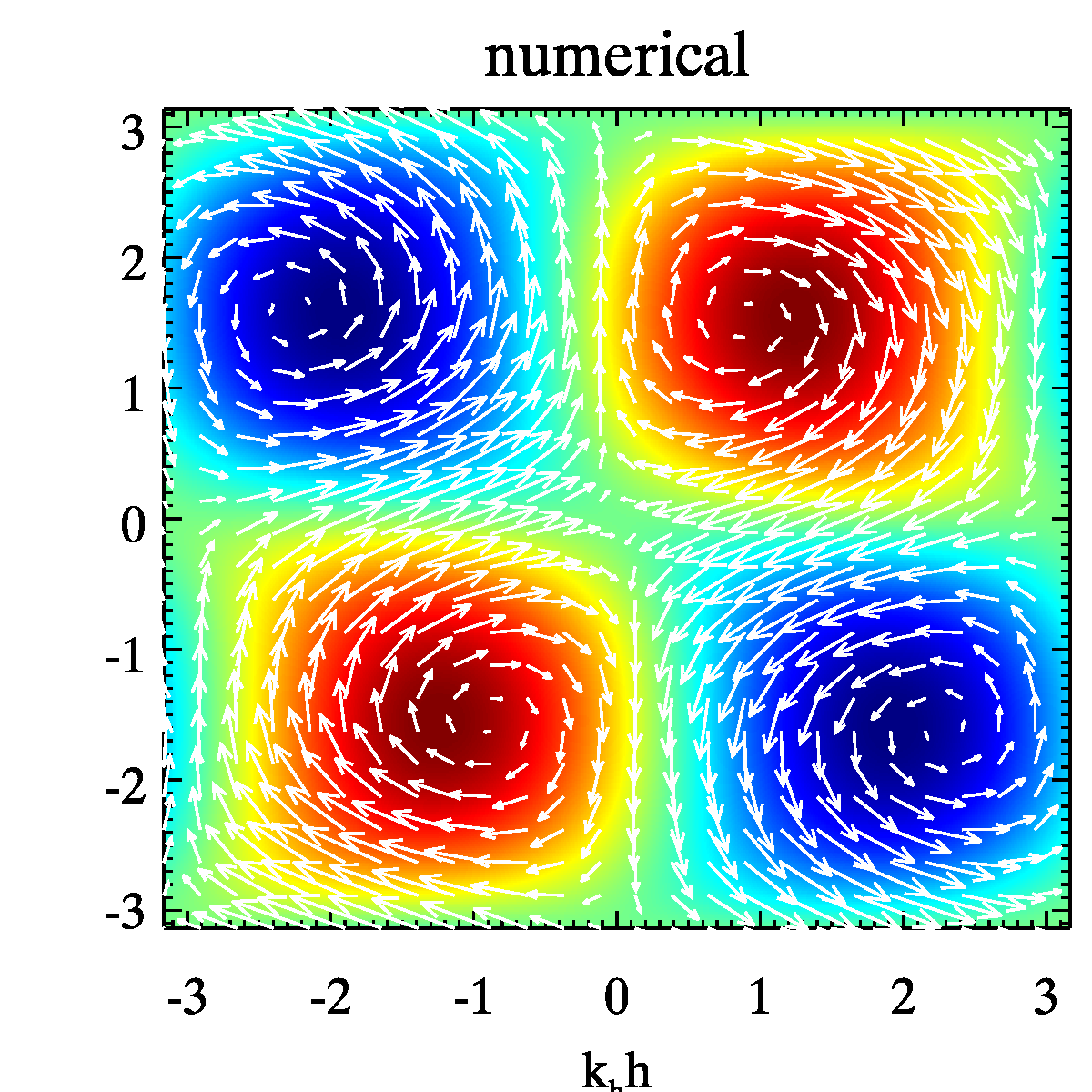}
\caption{
Comparison of 2D KH modes resulting from theoretical analysis (left
panel) and a numerical simulation with PLUTO (right panel), at $t=6{/(KV_0)}$.
The white arrows represent velocity vectors, overlaid on the vorticity
contours (color-map), associated with the KH instability, i.e., the
sinusoidal background shear profile has been subtracted.  The plots
show almost identical flow features, with similar peaks in the
vorticity and offset centers of the circulating velocity vectors.
}
\label{kh2danalytical_simcomparison_modes}
\end{center}
\end{figure}

\begin{figure*}[thbp]
\begin{center}
\includegraphics[width=0.99\textwidth]{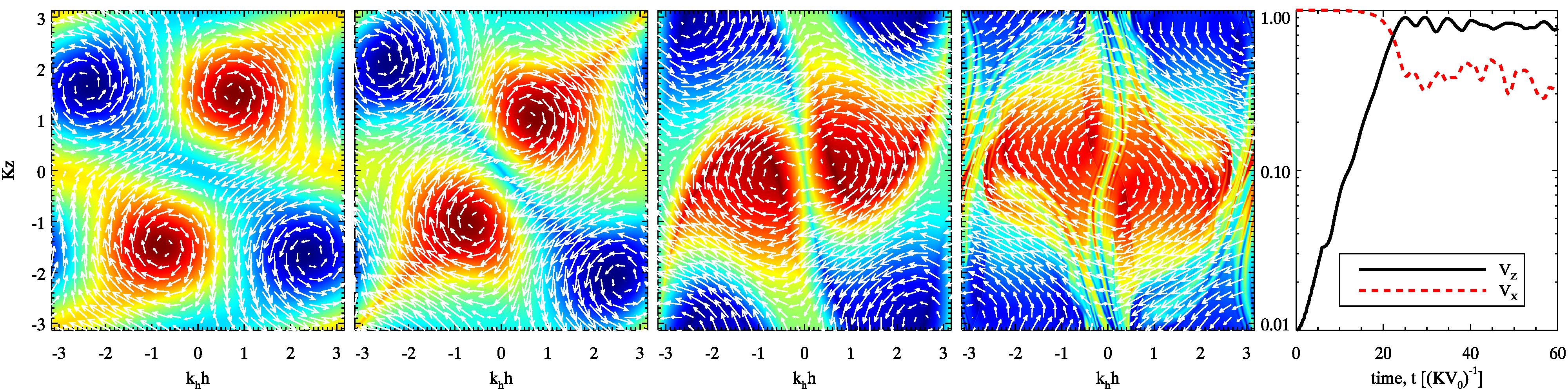}
\caption{
The first four panels illustrate the temporal evolution of the vorticity (colormap) and velocity field 
(white arrows) of the background-subtracted Kelvin-Helmholtz simulation. The time frames 
correspond to $t=10,20,30,40 ~(KV_0)^{-1} $, from left to right. The left-most panel shows the 
exponential growth of $v_z$ and decay of $v_{x}$, which brings the flow closer to an isotropic state.
}
\label{fig:khnonlin}
\end{center}
\end{figure*}

\subsection{Kelvin-Helmholtz Modes on a Sinusoidal Velocity Field}
\label{KHsection} 

We begin by examining the development of KH modes that results from a
background flow with $\delta v_x =V_0 \sin(2\pi z/L_z)$.  This
velocity profile is akin to the velocity profile induced by an MRI
mode, except for the fact that its amplitude is time independent. 
Incidentally,  note that a time-independent MRI amplitude is usually invoked
when calculating secondary instabilities affecting an MRI mode
\citep{Goodman:1994dd,Pessah:2010ic,Latter:2009br}. 
The argument
behind this approximation is that, provided that the amplitude of the
MRI is large enough, the growth rate of the secondary instability
should be much larger than the MRI growth rate.  This preliminary
exercise is useful for two reasons.  It provides an independent check
for the semi-analytical results predicting the parasitic growth rates
\citep{Goodman:1994dd} and mode structure \citep{Pessah:2010ic}. It
allows us to get acquainted with the type of mode structure that
emerges as a natural consequence of the action of parasitic modes in
3D shearing-box simulations of the MRI discussed below.

We solve the set of equations describing a
hydrodynamic flow starting from the initial conditions
\begin{align}
\bb{v}(x,z)=  \left( \begin{array}{c}  V_0 \sin(Kz) \\
0 \\
\epsilon V_0 \sin( k_h x_h ) \end{array} \right) \,,
\end{align}
 where $\epsilon = 10^{-2}$.
The stability of this flow can be analyzed semi-analytically as
described in \citep{Goodman:1994dd,Pessah:2010ic}.  The flow is
unstable with a dimensionless growth rate $s/KV_0$ that depends only
on $k_h/K$, i.e., the ratio of the wavenumbers associated with the
periodic background shear flow and the excited perturbation. This
growth rate attains a maximum at $k_h/K\simeq 0.6$ and vanishes beyond
$k_h/K = 1$.  This is illustrated in Figure 
\ref{kh2danalytical_simcomparison}, where we compare the results
obtained with linear theory and numerical codes.  In order to gauge
the importance of compressibility in the growth of the KH
instabilities, we performed two separate sets of simulations in
compressible and incompressible gas dynamics using the publicly
available codes PLUTO \citep{2012ApJS..198....7M} (where the equation
of state is isothermal) and SNOOPY \citep{Lesur:2009ey}.  As expected,
the results of the spectral code SNOOPY are closer to the analytical
results than the finite volume code, PLUTO, the difference is
accounted for by compressibility.

The early evolution of this sinusoidal, two-dimensional flow leads to
the development of sheets of cellular vortices of alternating
polarities both in radius and height, as shown in Figure
\ref{kh2danalytical_simcomparison_modes}.  The agreement between the
analytical and numerical results in the linear regime is also
excellent in terms of the structures formed.  The extrema of the
vorticity distribution is due to a combination of circulation and
shear vorticity, so the peak of the color map does not coincide with
the point around which the fluid seems to be circulating around.  
Figure  \ref{fig:khnonlin} shows the evolution of the instability into
the non-linear regime, where the vortices interact, grown and merge. 

The saturated state, depicted in the right-most panel of Figure~\ref{fig:khnonlin},
consists of episodic fluctuations in the vertical component of the velocity, 
which leads to a decreases of $v_z^2$ to about 50\% of its initial value.
The radial component of the velocity also shows episodic fluctuations,
which leads $v_x^2$ to increase to 80\% of the initial kinetic energy
(Figure~\ref{fig:khnonlin}). At later times, shock waves form, fine
sheared structures appear, and kinetic energy is dissipated into
thermal energy.  Note that while the initial shear flow is aligned in
the $x$-direction, the KH instability increases the flow speed in the
transverse, $z-$direction, until it reaches the same order of
magnitude, so that initially anisotropic flow becomes partially
isotropized.

\section{Development and Saturation of the MRI}
\label{latelinearMRI} 

\subsection{Beyond Semi-analytical Approximations }

In the standard picture, parasitic instabilities are thought to feed
off the exponentially growing background provided by the primary MRI
mode, which corresponds to an exact solution of the MHD equations
provided that compressibility is unimportant. In ideal MHD, the growth
rate of these secondary instabilities is expected to be proportional
to the amplitude of the MRI mode, and thus the growth of the parasites
is super-exponential. Therefore, the energy
obtained by the secondary instabilities can become comparable to the
energy in the MRI mode from which it feeds very fast.  Because the energy going
into secondary instabilities is envisioned to be directly supplied by the primary mode,
the fast growth of the secondary could potentially prevent the MRI
from continuing growing, halting its exponential growth and leading
thus to its saturation.

This picture assumes that the amplitude of the MRI is large enough so
that several approximations can be invoked. Perhaps more importantly,
it also assumes that the dynamics of the secondary perturbations is insensitive to the
shearing background and that the MRI grows uninhibited.  The first
approximation implies that the parasitic perturbations can be
described by a time-independent wavevector $\bb{k}_{\rm h} = (k_x,
k_y)$, which characterizes the wavelength of the perturbations in
horizontal planes, i.e., perpendicular to $z-$direction. However,
because of advection by the background shear flow, the wavevectors of
non-axisymmetric perturbations will evolve in time according to
$k_x(t) = k_{x} - q\Omega t \, k_y$.  The second approximation
prevents any dynamic coupling between the primary mode and the
secondary parasites. This approximation becomes increasingly worse as
the amplitude of the parasite grows larger.   In order to understand
the role of secondary instabilities, and how they mediate the transfer
of power from axisymmetric MRI modes into non-axisymmetric structures,
we set up a tailored numerical experiment which is not limited by the
assumptions usually invoked in the semi-analytical studies described
above.  For the sake of simplicity, we focus on following the
dynamical evolution of the fastest growing MRI mode.

\begin{figure}[t]
\begin{center}
\includegraphics[width=0.95 \columnwidth]{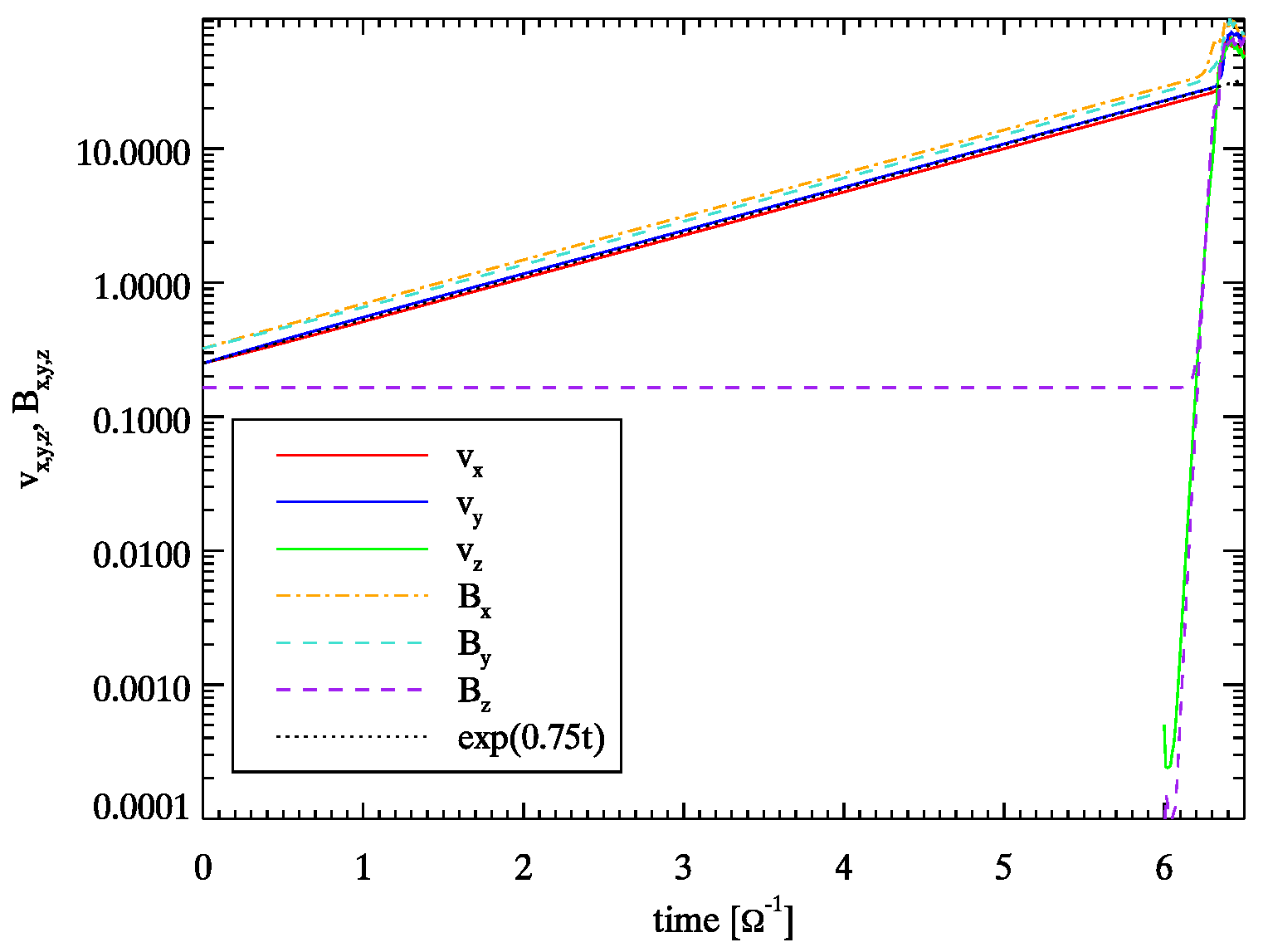}
\caption{
Time evolution of integrated quantities $B_{x,y,z}$ and $v_{x,y,z}$
plotted with a fit to the MRI growth rate of $0.75\Omega$. 
The field components $v_{x,y}$ and $B_{x,y}$ grow exponentially, while
$B_z$ remains constant until $t=6/\Omega$, when $v_z$
is perturbed and then grows super-exponentially.  
}
\label{fig:incompressiblemrigrowthrates}
\end{center}
\end{figure}

\subsection{Tailored Numerical Setup}

The incompressible simulations are carried out using the SNOOPY
spectral code \citep{Lesur:2009ey}, which solves the MHD equations
in the constant density regime and incorporates a finite physical
resistivity and viscosity.  The resistivity is such that $\Lambda_\eta
\equiv \bar{v}_{{\rm A} z}^2 / \eta \Omega = 10^4$, and the magnetic 
Prandtl number is unity. For practical purposes, this is rather close 
to ideal MHD and corresponds to the regime where the 
secondary Kelvin-Helmholtz mode is expected to dominate.  

We set up the numerical experiment as follows. In order to maximize
the numerical resolution, we set the vertical extent of the simulation
domain to be equal to the most unstable MRI wavelength, so that $L_z =
 2\pi K_{\rm max} =\sqrt{15/16}\, \bar{v}_{{\rm A}, z}/\Omega$. This
motivates choosing $L_z$ and $\Omega^{-1}$ as the units of length and
time.  All the results in the paper correspond to a simulation
with a resolution of 256x256x128 cells, which was carried out for $260\,\Omega^{-1}$. 
The only exception are the results shown in Fig \ref{fig:incompressiblemrigrowthrates400}, 
where we used the lower resolution of 128x128x64 cells, in order to 
run the simulation for $400\,\Omega^{-1}$.

We initialize the simulation by exciting the fastest growing MRI mode at $t=0$, i.e.,
\begin{align}
\bb{v}(x,y,z, t=0)=    \delta v \left( \begin{array}{c}
  \sin(K z) \\
  \sin(K z)  \\
0 \end{array} \right) \,,
\label{eq:IC_MRI_V}
\end{align}
\begin{align}
\bb{B}(x,y,z, t=0)= \delta B  \left( \begin{array}{c}
 \,\,\,\,\cos (K z) \\
-    \cos (K z) \\
0  \end{array} \right) \,,
\label{eq:IC_MRI_B}
\end{align}
where $\delta v =10^{-3} L_z \Omega$ and $\delta B/\sqrt{4\pi\rho} = \sqrt{5/3} \, \delta v$.
As expected, this initial perturbations grows exponentially at a rate 
$\Gamma_{\rm max} = 3/4\Omega$ (Figure~\ref{fig:incompressiblemrigrowthrates}) 
with the mode structure predicted by linear theory. In incompressible 
dynamics this exponential growth can continue for several orbits
unless the fluid is perturbed.

\begin{figure}[t]
\begin{center}
\includegraphics[width=0.45 \textwidth]{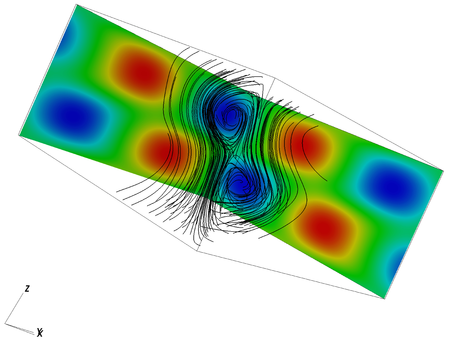}
\caption{
Vorticity (colormap) with exponentially growing MRI background subtracted 
and projected into the $x_p$  direction  at $t=6.05~\Omega^{-1}$ . 
The slices are in the $x,z$ and $y,z$ planes.
The black solid lines show the magnetic field.
}
\label{3dprojectedvorticity}
\end{center}
\end{figure} 

\begin{figure*}[t]
\begin{center}
\includegraphics[width=0.99 \textwidth]{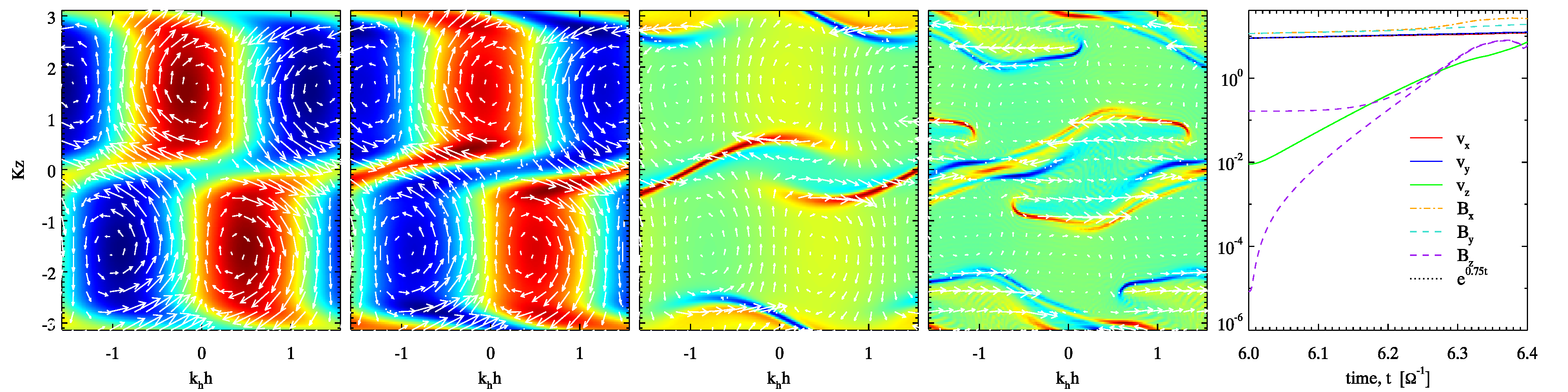}
\caption{
The first four panels illustrate the temporal evolution of the velocity field 
of the parasitic mode (white arrows) projected into the $x_h$ direction 
(where the MRI-induced shear is strongest) after subtracting the exponentially 
growing MRI mode. The color maps represent the associated vorticity contours.
The time frames correspond to  $t=6.15,6.2,6.3,6.4 ~ \Omega^{-1}$, from left to right. 
The left-most panel shows a zoom-in on the growth of the components
$v_{x,y,z}$ and $B_{x,y,z}$.
}
\label{2dcutof3dprojectedvorticity}
\end{center}
\end{figure*}

In order to break down the exponential growth of the MRI, we seed a 
perturbation with a horizontal wavevector given by $\bb{k}_{\rm h} = (1,1,0)$
\footnote{ Note that the horizontal wavenumber $k_{\rm h}/K \simeq 0.7$ 
is chosen to ensure compatibility of the periodic boundaries in horizontal planes, 
whereas the fastest growing parasite is $k_{\rm h}/K \simeq 0.6$, 
according to semi-analytical studies.} . Note that this wavevector is chosen in the direction where the
shear induced by the MRI is fastest, i.e., at 45 degrees with respect to the radial direction.
Because this mode is non-axisymmetric, its excitation requires some care for the following 
reason. If a non-axisymmetric perturbation is excited at time $t=0$, 
since the radial wavenumber has a linear dependence on time, it 
reaches the Nyquist wavenumber in $t_{alias}=\log_2 (N_{x,g}) ~\Omega^{-1}$ 
(where $N_{x,g}$ is the number of grid cells in the x-direction), 
where it is aliased to $k_x=1$ before it can grow significantly 
\citep{Lyra:2011il}. To circumvent this, we excite the perturbation
\begin{align}
\delta \bb{v}(x,y,z, t=6) =  \epsilon \, \delta v \sin( \bb{k}_{\rm h} \cdot \bb{x}) \,\check{\bb{z}} \,,
\label{eq:IC_PI_V}
\end{align}
with $\epsilon=10^{-4}$ at time $t=6\Omega^{-1}$.
This allows the perturbation to seed the parasitic mode,
avoiding the effects of aliasing due to finite grid resolution.  This
secondary instability grows super-exponentially
(Figure~\ref{fig:incompressiblemrigrowthrates}), exhibiting the
characteristics expected of a KH instability feeding off the velocity
shear induced by the primary MRI mode
(see Figure~\ref{3dprojectedvorticity} and  \citealt{Pessah:2010ic}).  
The subsequent evolution of this
three-dimensional, non-axisymmetric perturbation, in the plane 
parallel to $\bb{k}_{\rm h}$, i.e.,
where the MRI-induced velocity shear is largest, is
shown in Figure~\ref{2dcutof3dprojectedvorticity}.  Note that the
rightmost panel there shows a close-up of
Figure~\ref{fig:incompressiblemrigrowthrates} at the time where the
amplitude of the secondary instability approaches that of the MRI.

In order to analyze the dynamics in detail, it is useful to divide the
evolution of the flow into three stages: {\it i}) an exponential
growth phase driven by the MRI in which the amplitude of the fast
growing parasite is too small to modify the primary mode.  {\it ii}) A
transition stage in which the energy of the parasite is comparable to
the energy of the primary MRI mode, which starts to depart
significantly from the initial exact solution.  {\it iii}) A turbulent
state in which all flow variables fluctuate in time.  In what follows,
we describe briefly each of this stages.

\begin{figure*}[t]
\begin{center}
\includegraphics[width=0.99 \textwidth]{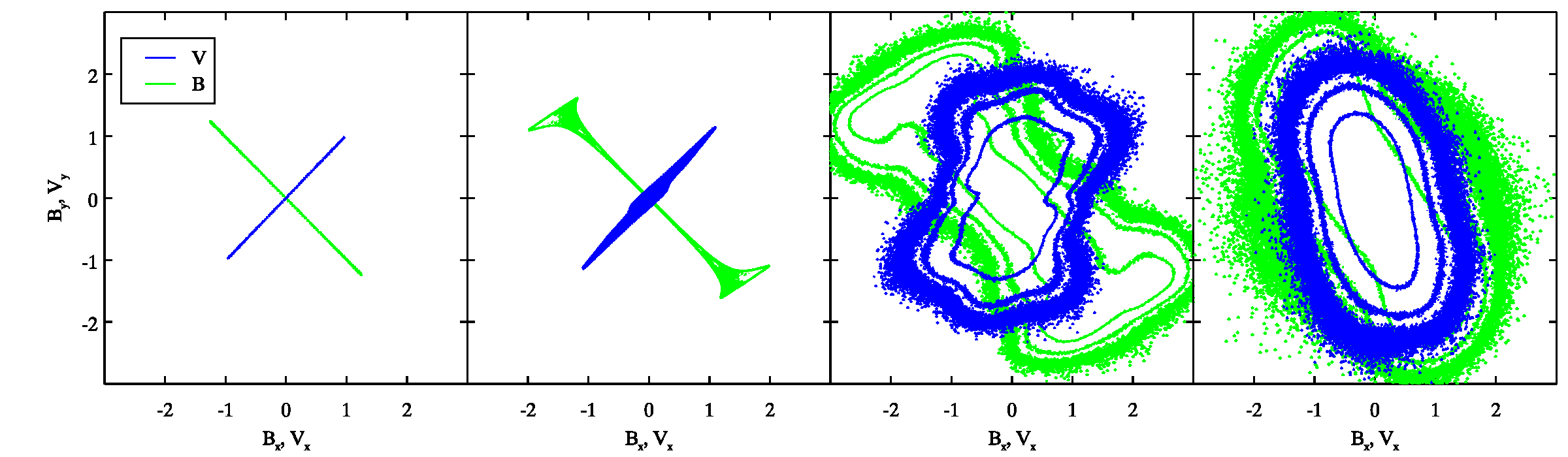}
\caption{
Scatter diagrams of horizontal magnetic  and velocity fluctuations 
$B_x, B_y$ (green contour) and $v_x, v_y$ (blue contour) at times 
$t=6.0,6.5,6.6,15 ~ \Omega^{-1}$, from left to right.
At time $t=6.0$, the magnetic and velocity 
modes are linear, orthogonal, and highly anisotropic. 
During the growth of the parasite mode, ($t=6.5\Omega^{-1}$), the magnetic 
energy departs from the linear state at high values of magnetic field, and begins 
to partially isotropize. 
In the saturated state ($t=15\Omega^{-1}$), the flow is still anisotropic.
We use the projection introduced in \citet{Pessah:2008bf}, see, e.g., their Figure~2.
}
\label{fig:phasespacebxby}
\end{center}
\end{figure*}

\subsection{MRI Exponential Growth in the Linear Regime}

The initial conditions in equations (\ref{eq:IC_MRI_V}) and
(\ref{eq:IC_MRI_B}) lead to exponential growth of the exact solution
which grows for several $~\Omega^{-1}$ timescales. The amplitude of this perturbation
grows by more than two orders of magnitude retaining the initial MRI
mode structure in which the sinusoidal velocity field, with $\delta
v_x = \delta v_y$, is orthogonal to the co-sinusoidal magnetic field
perturbations, with $\delta B_x = \delta B_y$.  This is illustrated in
the left-most panel in Figure~\ref{fig:phasespacebxby}, which shows
the magnitude of the radial and azimuthal components of the velocity
and magnetic field \citep{Pessah:2008bf} as a scatter plot \citep[see, e.g.,][]{2008A&A...487....1B}. 

In the incompressible regime under consideration, the exact MRI mode
does not lead to velocity or magnetic field perturbations in the
vertical direction. Therefore keeping track of $\delta v_z$ and
$\delta B_z$ allows us to quantify how the evolution of the flow
departs from the exact MRI mode excited at $t=0$. Figure~ 
\ref{fig:incompressiblemrigrowthrates} illustrates that the
vertical component of the magnetic field, that gives rise to the MRI,
remains constant until the secondary perturbation (\ref{eq:IC_PI_V})
is excited at $t=6\Omega^{-1}$. The initial perturbation seeding $\delta v_z$ 
grows super-exponentially together with $\delta B_z$, affecting
all other field components, which start to depart significantly 
from the MRI behavior predicted by linear theory soon after. This 
is also evident from the middle and right-most panels in Figure~\ref{fig:phasespacebxby}.

The shear resulting from the exponentially growing sinusoidal
velocity field that results from the MRI is maximum along the 
direction $x=y$, which is the direction in which we have chosen $\bb{k}_{\rm h}$.  
This plays an important role in the three-dimensional, non-axisymmetric 
nature of the parasitic modes.
The evolution of the parasitic mode is illustrated in Figure~
\ref{2dcutof3dprojectedvorticity}.  The first four panels, from left
to right, show the projected component of the vorticity into the
$(x_h,z)$ plane, with $x_{\rm h} = \bb{x} \cdot \bb{k}_{\rm h}$. 
The initial vorticity perturbation, which is of the form
$\nabla \times \bb{v}_{\rm h} \propto \cos ( \bb{k}_{\rm h} \cdot \bb{x})  \, \hat{\bb{k}}_{\rm p}$,
with  $\hat{\bb{k}}_{\rm p} = \hat{\bb{k}}_z \times \hat{\bb{k}}_{\rm h}$,
gives rise to a sheet of cellular vortices, as predicted in
\cite{Pessah:2010ic}.  The first and second panels, from left to right,
show that the cellular vortices begin to interact in the nonlinear
stage and vorticity concentrates near the boundary of the vortex
sheets, at $Kz=0,\pm 0.5$. 
The sheets of vortices can be seen in the 3D slices of vorticity and
velocity field lines projected into the $(x_{\rm h}, z)$ plane (see
Figure \ref{3dprojectedvorticity}), where the familiar parasite structure exhibits 
extrema visible at $z=-.25, 0.25$.

\subsection{Takeover by Parasitic Instabilities}

The vorticity plots in the third and fourth panel in Figure~\ref{2dcutof3dprojectedvorticity} 
illustrate how the interface undulations roll over into the classical cat's eye
associated with Kelvin-Helmholtz vortices, which appears between the
two shearing layers, similar to those seen in
\citet{Ryu:1996ha,Jones:1997it,Malagoli:1996um}.  At this time
(6.3-6.4 $~\Omega^{-1}$) the peak value of the magnetic field $B_z$ exceeds
the peak value of the vertical velocity component, $v_z$.  The radial
component of the magnetic field has also started to depart from its linear
growth and exceeds the MRI growth rate, as can be seen from Figure~\ref{fig:incompressiblemrigrowthrates}.
As a result of the super-exponential growth of the KH instability, the
flow becomes completely disrupted and transitions rapidly into a turbulent state.

It is instructive to compare the evolution of the secondary KH instabilities
that feed off the exponentially growing sinusoidal shear provided by the MRI and the
constant, imposed sinusoidal shear addressed in Section \ref{KHsection}.
Because the grow of the MRI mode induces magnetic field perturbations that are 
orthogonal to the velocity fluctuations, the nature of the motions in the plane
$(x_h,z)$ is essentially hydrodynamic. Therefore, it is not surprising that the
perturbations (\ref{eq:IC_PI_V}) lead to the excitation of
Kelvin-Helmholtz instabilities similar to the ones described in
Section \ref{KHsection}. However, because the amplitude of the
velocity shear from which the secondary instability feeds off is
growing exponentially in time, their late temporal evolution exhibits
significant differences with respect to the KH instabilities triggered
by the constant amplitude sinusoidal profile (compare
Figures~\ref{fig:khnonlin} and \ref{2dcutof3dprojectedvorticity}).

In the linear stage of the MRI, the flow is strongly anisotropic, as the vertical
components of the magnetic field and velocity are greatly
exceeded by their radial and axial counterparts.  
The departure from the linear state can be seen in the scatter plots 
of $B_{x,y}$ and $v_{x,y}$ (see Figure~\ref{fig:phasespacebxby}), the second panel shows the
departure from the linear mode at $t=6.5 ~\Omega^{-1}$, shortly after the
perturbation was excited.  The magnetic field components $B_x$ and $B_y$ no longer increase
exponentially, but instead a parasite mode starts to take over. 
The $B_x, B_y$ structures, which grow
at right angles to the initial MRI flow, at $(B_x,B_y)=(-2,1)$ and
$(2,-1)$ in the second panel, then correspond to the growth of
parasitic modes.  
At saturation time,
all three components of the velocity and magnetic fields are
approaching the same magnitude.  In essence, the growth of parasite 
modes has a strong effect on the anisotropy of the flow, reducing it and 
attempting to return to isotropy, as it can also be seen from 
in Figure~\ref{fig:phasespacebxby}.

The MRI mode starts to saturate as energy is drained from the radial
and azimuthal components of velocity and magnetic fields and fed into
the vertical components of velocity and magnetic fields.  Thus a
reduction in the strong anisotropy, from $v_x \sim v_y \gg v_z$, to
$v_x \sim v_y \sim v_z$, appears to accompany the parasite modal growth.

\subsection{Quasi-Periodic Turbulent State}

\begin{figure}[b]
\begin{center}
\includegraphics[width=0.99 \columnwidth]{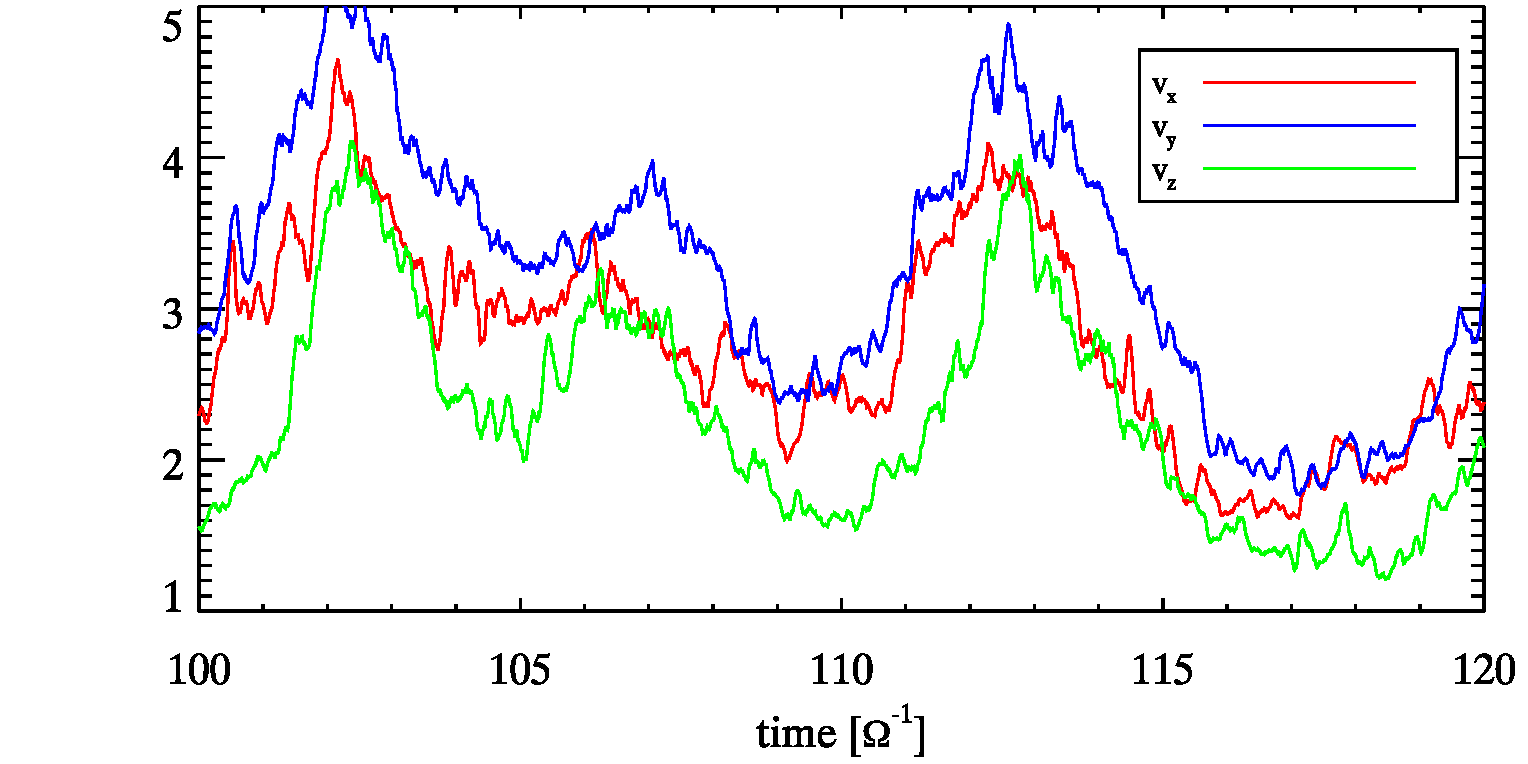}
\includegraphics[width=0.99 \columnwidth]{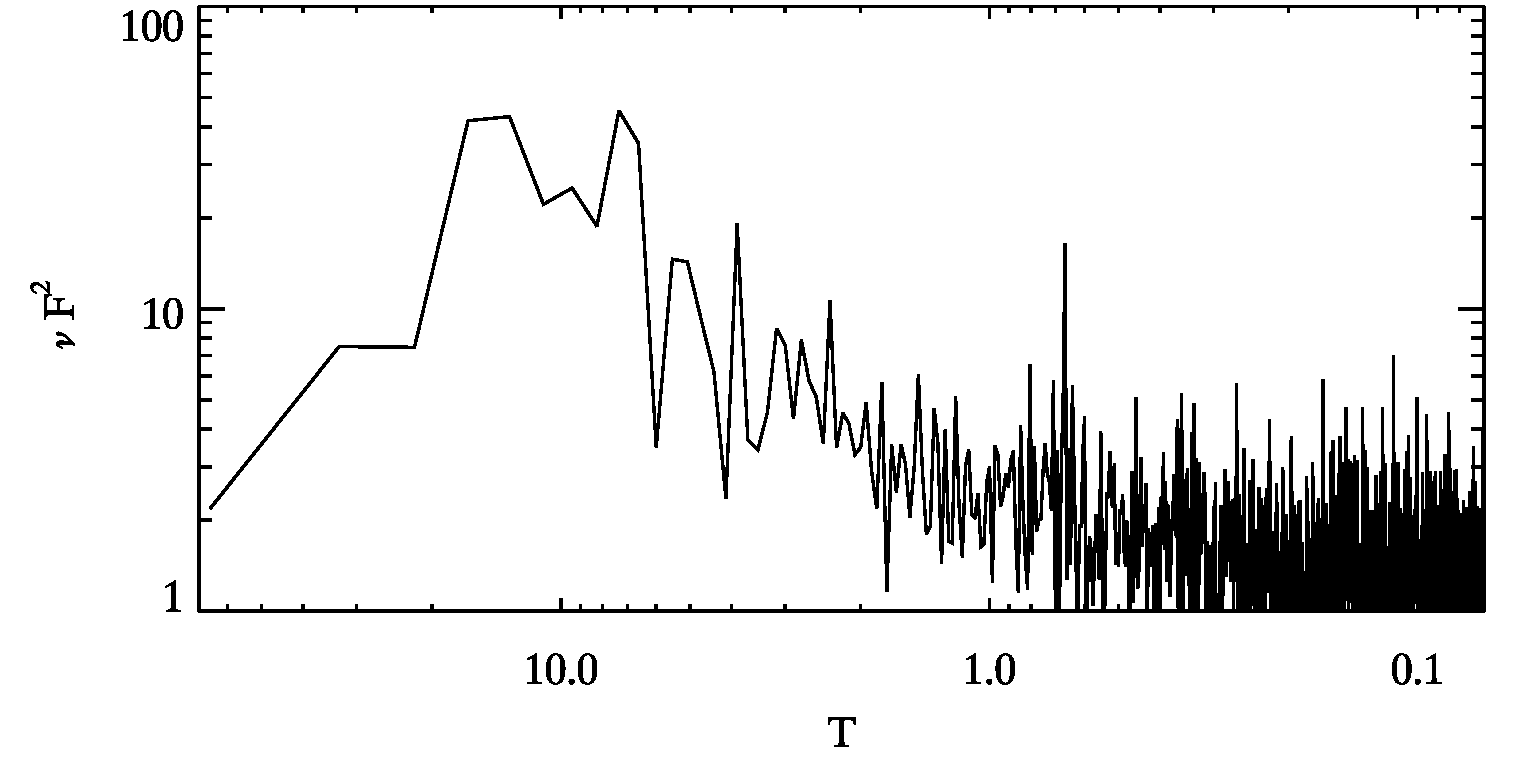}
\caption{
Upper panel: Time evolution of  the radial and vertical velocity components, 
$v_{x}$ and  $v_{z}$, in the range $t=100-120~\Omega^{-1}$. 
The radial component $v_x$ always starts growing before the vertical component
$v_z$,  which then grows at a faster rate, until both variables peak at the 
same value and drop to a local minimum. 
Lower panel:  Temporal power spectrum resulting from averaging the square 
of the Fourier transforms corresponding to four independent samples of the 
time series of the radial component of the velocity.  
}
\label{fig:incompressiblemrigrowthrates400}
\end{center}
\end{figure}

After the parasite mode has grown to its maximum value at $t= 6.6~\Omega^{-1}$,
and several $~\Omega^{-1}$ have elapsed, a state of fully developed,
anisotropic turbulence is reached. The analysis of the MRI-driven 
turbulent state has been the subject of a large body of literature.
Here, we point out that the turbulent regime observed in our numerical
experiment is characterized by sequences of peaks in the $v_x$ component
followed  by peaks in $v_z$.  These quasi-periodic oscillations are evident in 
Figure~\ref{fig:incompressiblemrigrowthrates400}.  The radial component $v_x$
always increases more slowly, followed by a rapid increase in the vertical component
$v_z$. As both components attain their respective local maxima, they
decrease in tandem and at the same rate.  Interestingly, this echoes the behavior 
seen in the linear stage, where the growth in $v_x$ and $v_{y}$ (driven by the MRI 
mode) triggers the growth in $v_z$ (resulting from the excitation of parasite modes) 
which then prevent the amplitude of the MRI from growing further.

In order to quantify this periodicity, we analyze the time series
provided by $v_x(t)$.  We consider a total of $350 ~\Omega^{-1}$, 
discarding the initial 50 $~\Omega^{-1}$ in which the simulation 
is growing exponentially and then transitioning towards fully turbulent flow.
We use Bartlett's method of estimating power
spectra via averaging periodograms \citep{1948Natur.161..686B} in
order to reduce their variance. This is achieved by splitting the
the time series for $v_x(50<t<300)$ in four independent samples,
calculating the Fourier amplitudes of these time series, and averaging
their square for each frequency (see
Figure~\ref{fig:incompressiblemrigrowthrates400}).  The bulk of the
power is concentrated between 9 and $13~\Omega^{-1}$.  A spike is also seen
at two-thirds of a period, which corresponds to one shearing time.

\begin{figure*}[ht]
\begin{center}
\includegraphics[width=0.89 \textwidth]{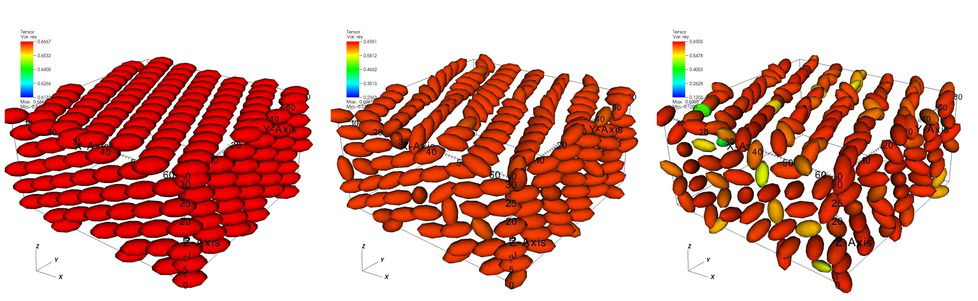}
\caption{
Rendering of the temporal evolution of the Reynolds stress anisotropy tensor,
$\bb{\mathcal{R}}$, at $t=6.2 ~\Omega^{-1}$, $t=6.3 ~\Omega^{-1}$
and $t=6.5 ~\Omega^{-1}$, from left to right.
The principal axes of the ellipsoids are given by the three eigenvectors
and they are aligned with the direction of the principal eigenvector.
The colors indicate the magnitude of the maximum eigenvalue. In the linear 
phase of the MRI, and immediately before saturation, $\bb{\mathcal{R}}$ is 
constant and aligned everywhere in the $x=y$ direction, i.e., 
the direction where the MRI-induced shear 
is strongest.  The second panel shows $\bb{\mathcal{R}}$
immediately after the onset of the parasitic modes
The parasites have grown and reduced the
principal stress and scattered its direction in different regions of the domain.  The
third panel shows $\bb{\mathcal{R}}$ well into the saturated phase.
}
\label{maxtensor}
\end{center}
\end{figure*}

In the saturated state, the flow is characterized by highly anisotropic
turbulence, which exhibits quasi-episodic bursts.
\citet{2008A&A...487....1B} attributed these presence (absence) of 
these bursts to the absence (presence) or  of parasite modes.  
In the linear stage, it has been possible
to directly subtract the background MRI exact solution, in order to
identify signatures and compare with those predicted in
semi-analytical calculations \citep{Pessah:2010ic}.  However, in the
turbulent state, the same approach becomes more difficult.  Exact
solutions are not easy to identify or subtract in a consistent way.
Therefore we pursue an alternative approach, in order to engage directly
with the fluctuations in the Reynolds and Maxwell stresses that are 
associated to angular momentum transport in the turbulent state.

\newpage

\section{Anisotropy of MRI turbulence}
\label{anisotropy}

In section \ref{latelinearMRI}, we presented numerical evidence of the
growth of parasitic instabilities responsible for the breakdown of the linear
exponential growth driven by the MRI.  Just as the standard
Kelvin-Helmholtz instability partially isotropizes the shear flow that
gives rises to it, the growth of the Kelvin-Helmholtz parasitic modes
feeding off the MRI background acts to suppress the anisotropy imposed
by the MRI-induced shear flow. It is plausible that, if parasitic modes
play a role in the saturated state as well as the late linear stage,
the anisotropy of the MRI-turbulent flow fluctuates in the 
saturated state as well. In this section, we analyze in detail 
how the anisotropy of MRI-driven turbulence varies with time.

The statistical nature of the turbulent state makes it necessary to
employ averages in order to quantify in a meaningful way the
properties of the flow.  The power spectra offers a useful tool to
understand what scales contribute to the process of interest.
Most statistical analyses of MRI-driven turbulence 
carried out so far rely either on averages along the axes in Fourier space
or spherical shell averages in Fourier space, and usually involve
time-averaging over periods that can span up to 50 orbits.
While taking cuts of Fourier space along the Cartesian axis is simple,
the results might provide a biased view of the underlying anisotropic 
power spectrum.  This is spatially true if one considers the plane spanned
by the directions along $k_x$ and $k_y$ where MRI- driven turbulence
is known to exhibit strong anisotropy.  Averaging over shells in Fourier space is a 
sensible approach in the case of isotropic turbulence. However, this is not the case 
for MRI-driven turbulence. Furthermore, it is clear from 
Figure~\ref{fig:incompressiblemrigrowthrates400} that the flow properties, 
including the degree of anisotropy, vary strongly on timescales of the order of 
$\sim10 ~\Omega^{-1}$. Averaging over longer periods of time could mask valuable 
information about the properties of the turbulence that we are aiming to understand.

\begin{figure*}[t]
\begin{center}
\includegraphics[width=0.49 \textwidth]{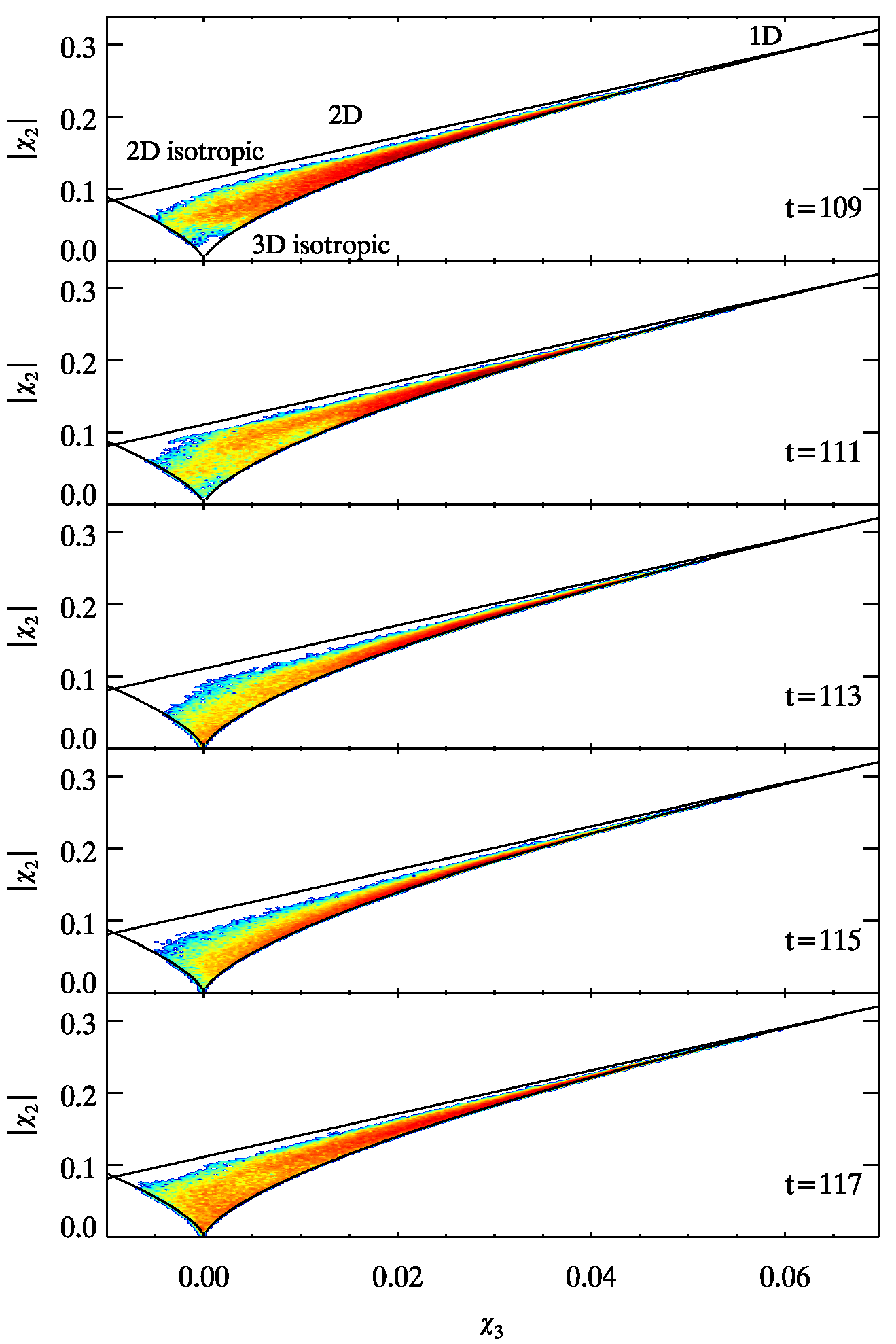}
\includegraphics[width=0.49 \textwidth]{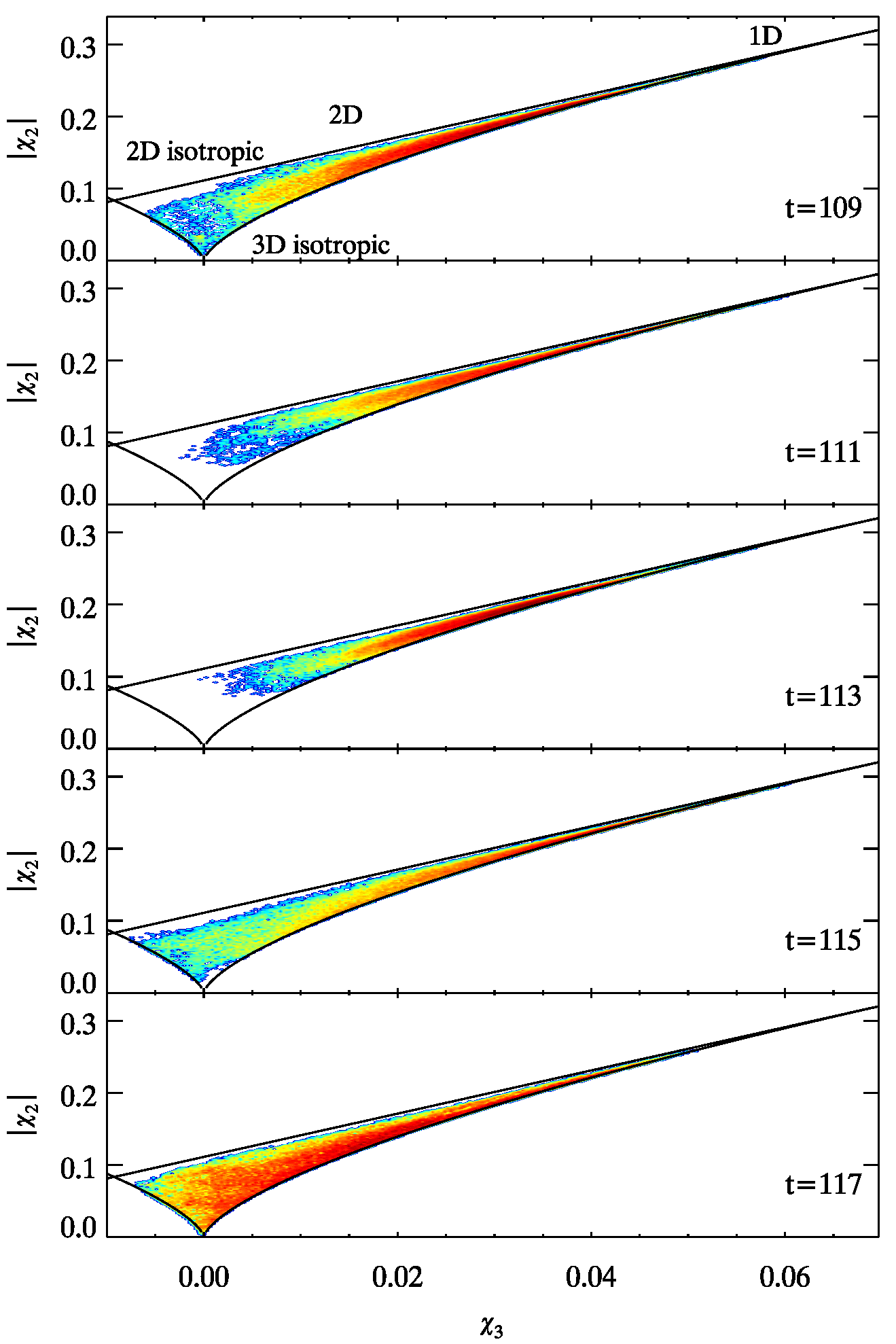}
\caption{
Temporal evolution of the invariant maps for the Reynolds (left column) and Maxwell  
(right column) stress  tensors at $t=40-48~\Omega^{-1}$ . The points show the second 
and third invariants, $\chi_2$ and $\chi_3$ respectively, of the Reynolds stress anisotropy 
tensor, for the saturated stage of the MRI.  The lines delineate the boundaries on the 
normalized invariants. The data points near the upper right vertices of the Lumley triangle 
are indicative of one-component turbulence. The data points close to the lower vertex have 
three equal eigenvalues, denoting isotropic three-dimensional turbulence, whereas the left 
vertex denotes two-component turbulence.  The color code denotes the number density of 
subdomains over which the averages were taken that have the same value the anisotropic 
invariants.
}
\label{lumleyv113}
\end{center}
\end{figure*}

\subsection{The Temporal Evolution of Invariant Tensor Properties}
\label{lumley}

In order to obtain a better understanding of the anisotropy of the flow, and its temporal
evolution, we employ the turbulent stress invariant analysis introduced by 
\citet{1977JFM....82..161L} in the context of hydrodynamics and apply it to study 
MRI-driven turbulence.

The Reynolds and Maxwell stress tensors are defined according to 
\begin{eqnarray}
R_{ij}&=& \rho \, \langle \delta v_i \, \delta v_j \rangle \,, 
\label{eq:reynolds_stress}\\
M_{ij}&=& \frac{\langle \delta B_i \, \delta B_j \rangle}{4\pi} \,, 
\label{eq:maxwell_stress}
\end{eqnarray}
where the brackets represent an appropriate spatial average.
Each of these real and symmetric tensors can be divided into a deviator and an isotropic part.
By normalizing the tensors using their trace, associated to the corresponding
energy densities, and then subtracting the isotropic component, we arrive at a 
single tensor quantity which parametrizes the deviation from isotropy.
The Reynolds stress anisotropy tensor \citep{1977JFM....82..161L} is defined as:
\begin{equation}
\mathcal{R}_{ij}=\frac{R_{ij} }{{\rm Tr}\,(R_{ij}) } - \frac{1}{3} \delta_{ij} \,,
\end{equation}
where ${\rm Tr}$ stands for the trace and $\delta_{ij}$ is the Kronecker delta. 
The Reynolds stress anisotropy tensor provides an 
important diagnostic and has been used extensively in numerical and experimental
studies of hydrodynamic turbulence \citep{Biferale:2005cwa}. Here, we extend this idea
in order to study of MRI-driven turbulence and define the Maxwell stress anisotropy tensor as
\begin{equation}
\mathcal{M}_{ij}=\frac{M_{ij} }{{\rm Tr}\,(M_{ij}) } - \frac{1}{3} \delta_{ij} \,.
\end{equation}

The evolution of the global distribution of anisotropy when parasitic
modes grow by feeding off the MRI can be quantified via
$\mathcal{R}_{ij}$ in equation (\ref{eq:reynolds_stress}) by averaging
over cubic volumes composed of eight cells. At a given time, this
procedure leads to a set of ellipsoidal glyph's as shown in
Figure~\ref{maxtensor} for three different time snapshots.  The
orientation and shape of each glyph is controlled by the eigenvectors
of $\bb{\mathcal{R}}$ and their corresponding eigenvalues. The latter,
defined in descending order as $\lambda_1,\lambda_2$, and $\lambda_3$,
correspond to the principal, medium and minor axes of the ellipsoidal
glyph. The direction of the glyph is aligned along the principal
stress eigenvector, i.e., the eigenvector with the largest associated eigenvalue.  
The left-most panel shows that all the tensors
are aligned in the direction along which the MRI-induced shear is
strongest, i.e., $x=y$, indicating that in the late linear stage of the MRI we have
highly organized and highly anisotropic flow. 
The middle panel shows $\bb{\mathcal{R}}$, after
the secondary perturbation has been excited and the parasites have
started to grow. The initially highly anisotropic flow has now been
partially isotropized and the principal stress eigenvectors of
$\bb{\mathcal{R}}$ are no longer parallel over the domain.  The
rightmost panel shows the partially isotropized field in saturation.
In this last panel, the system has not yet reached fully developed
turbulence but it has reached the closest approach to isotropy.

In order to probe how the anisotropy on larger scales varies as a function 
of time in a quantitive way, it is useful to employ invariant quantities. 
For a traceless tensor, $\lambda_1+\lambda_2+\lambda_3 = 0$, 
the degree of anisotropy can be characterized by calculating the two nonzero invariants 
\begin{align}
\chi^\mathcal{R}_2 &= - (\lambda_1^2+\lambda_2^2+\lambda_3^2) \,, \\
\chi^\mathcal{R}_3 &= \lambda_1 \, \lambda_2 \, \lambda_3 \,.
\end{align}
By plotting all the possible values of these two independent
invariants of the Reynolds stress in the plane spanned by
$(\chi_2,\chi_3)$, \cite{1977JFM....82..161L} found that every
realizable Reynolds stress is constrained to lie within three loci
which intersect at three vertices.  The enclosed deltoid, known as the
Lumley triangle, represents all possible values for
$\bb{\mathcal{R}}$.  The upper right vertex corresponds to the case of
$\lambda_1 > \lambda_2 , \lambda_3$, and the origin corresponds to
isotropy, $\lambda_1= \lambda_2 = \lambda_3$.  The upper left vertex
corresponds to two component turbulence, with  $\lambda_1 , \lambda_2 >
\lambda_3$. The vertical axis in these plots corresponds to the
deviation from isotropy, from zero (fully isotropic) through 0.083
(two-component isotropic) to 0.333 (one-component). The horizontal
axis in the plots correspond to the determinant, which distinguishes
between those cases which have the same magnitude of deviation but a
different sign.

\begin{figure*}[t]
\begin{center}
\includegraphics[width= \textwidth]{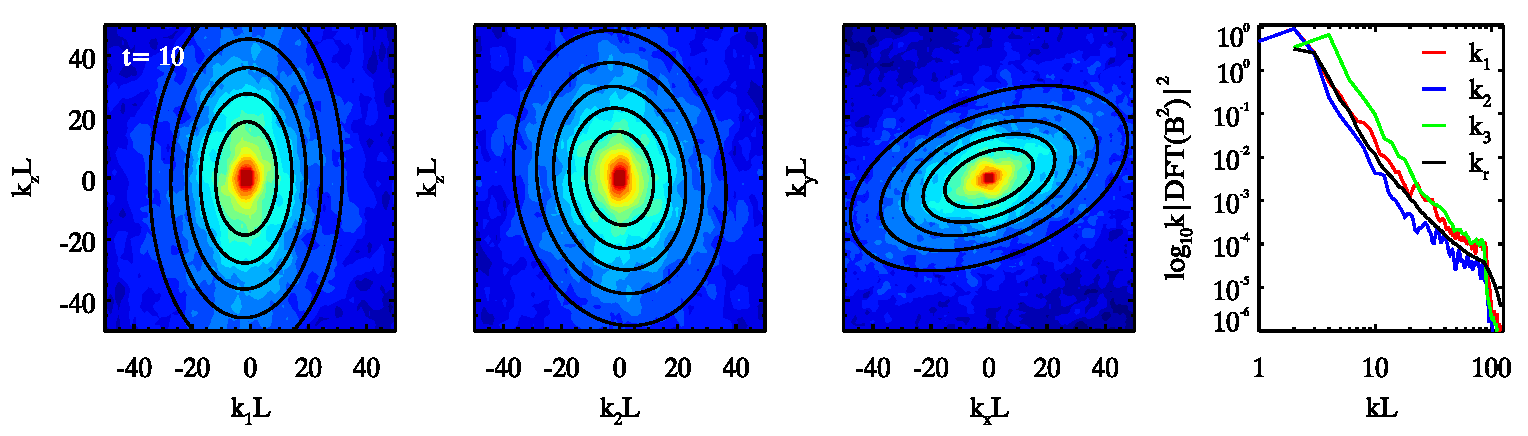}
\includegraphics[width= \textwidth]{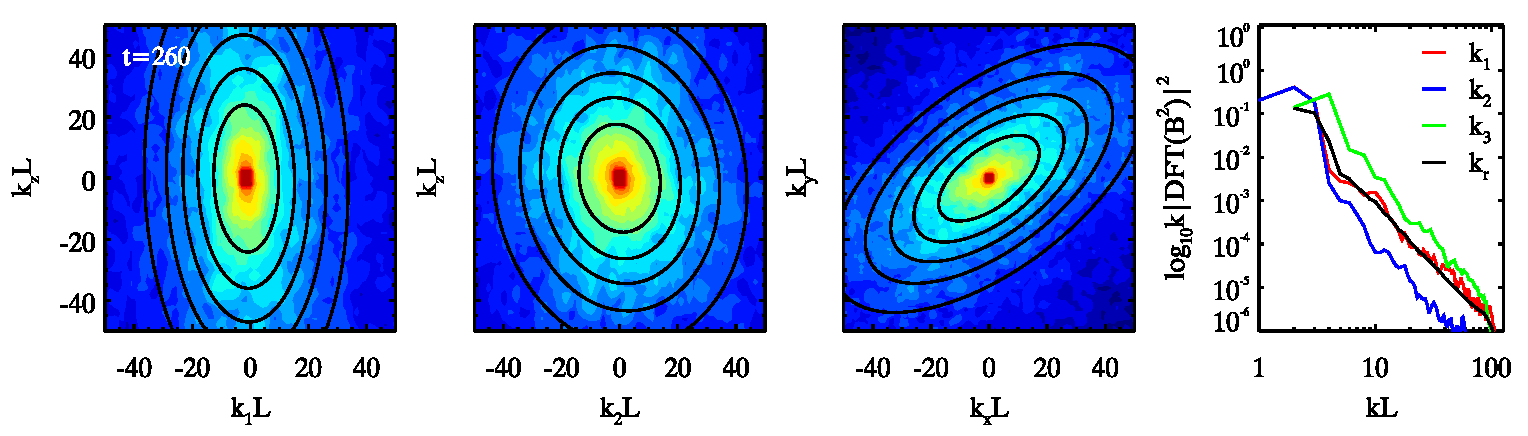}
\caption{
Analysis of the power spectral density of the magnetic energy density in the 
early stages of the nonlinear evolution of the MRI and the fully developed 
turbulent state. The upper and lower row correspond to $t=10~\Omega^{-1}$  
and $t=260~\Omega^{-1}$, respectively.
The first three panels in each row, from left to right, show slices of the 3D discrete 
Fourier Transform (DFT) of the magnetic energy density in the $(k_1,k_z)$,$(k_2,k_z)$ 
and $(k_x,k_y)$ planes, respectively. The data have been smoothed by using a box average filter over every two cells. The black contours in each panel are 2D gaussian elliptical fits to the 
smoothed distributions. The right-most column shows 1D cuts of the same data, compared with the spherical average as a function of the radial wave vector. 
It is evident that the turbulence exhibits a high degree of anisotropy. In particular, there is 
difference of almost two orders of magnitude in power between the 1D spectra along $kz$ and $k_2$.
}
\label{cutsof3dfft}
\end{center}
\end{figure*}

We provide a quantitative representation of the time evolution of
anisotropy on larger scales, by showing in Figure~\ref{lumleyv113} 
how the Lumley triangles associated with the
Reynolds and Maxwell stresses are populated as a function of time.
The plots are constructed by taking contours of 2D histogram of the
distribution of vertically averaged tensors.  
The left column shows the Reynolds stress tensor invariant map at five successive times. 
As time progresses the distribution of anisotropy fluctuates initially 
increasing between $109<t~\Omega^{-1}<113$ and later decreasing between $113<t~\Omega^{-1}<117$. 
The left column of Figure~\ref{lumleyv113} shows the normalized Reynolds stress tensor invariant map, 
with the second invariant in the vertical axis and the third invariant (determinant) in the horizontal axis.
The color map shows the density of points that have associated the pair of values $(\chi_2, \chi_3)$ 
when each stress tensor is averaged in the vertical direction.
At $t=119~\Omega^{-1}$, the stress tensors have a peak which is far from the origin. The highest density (shown in red) 
is offset from the origin which is consistent with anisotropic turbulence. 
The majority of points are distributed close to the 1D-3D locus, with almost no points on the 1D-2D locus.
As time progress, at $t=111~\Omega^{-1}$ and $t=113~\Omega^{-1}$ the 
distribution of anisotropy shifts along the 1D-3D locus, further from isotropy (i.e., the origin).
At $t=113~\Omega^{-1}$, the peak of anisotropy is reached and almost no power present near the origin.
At $t=115~\Omega^{-1}$, the distribution of invariants shift back towards the origin, indicating a return towards isotropy. 
The peak of the distribution remains far from the origin, however.
At $t=117~\Omega^{-1}$, the distribution shifts even further towards the origin.

This effect is even more pronounced in the Maxwell stress tensor (right column of Figure~\ref{lumleyv113}) , 
which shows a strong increase in anisotropy between $109<t~\Omega^{-1}<113$, which then reverses between 
$113<t~\Omega^{-1}<117$. The examination of other timeframes (not reproduced here) reveals that this
pattern of increase and decrease in anisotropy is quasi-periodic, in agreement with Figure~\ref{fig:incompressiblemrigrowthrates400}.

\begin{figure*}[t]
\begin{center}
\includegraphics[width= \textwidth]{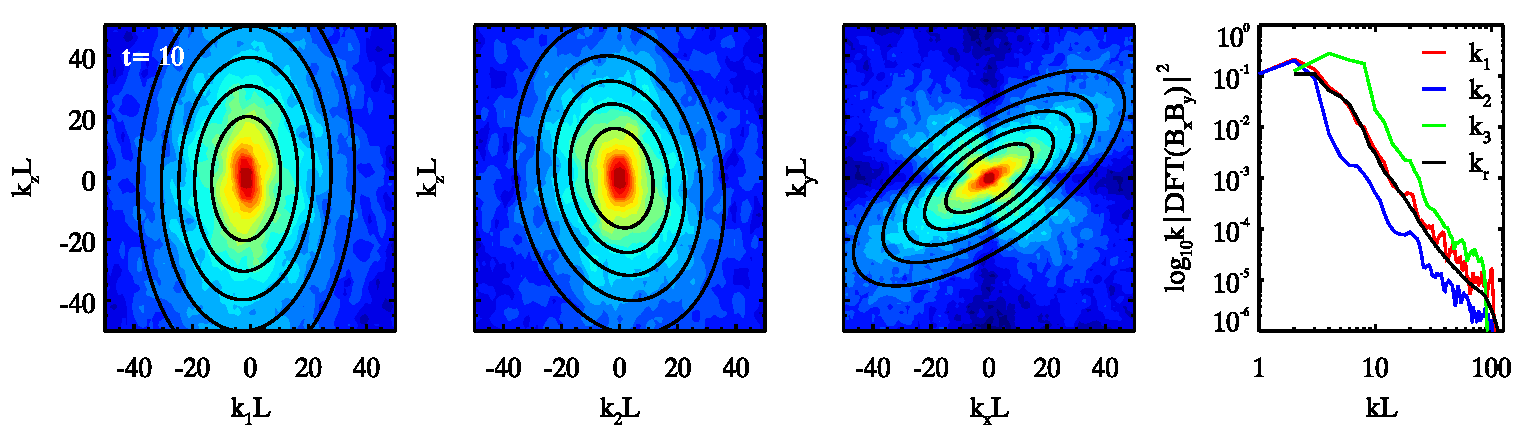}
\includegraphics[width= \textwidth]{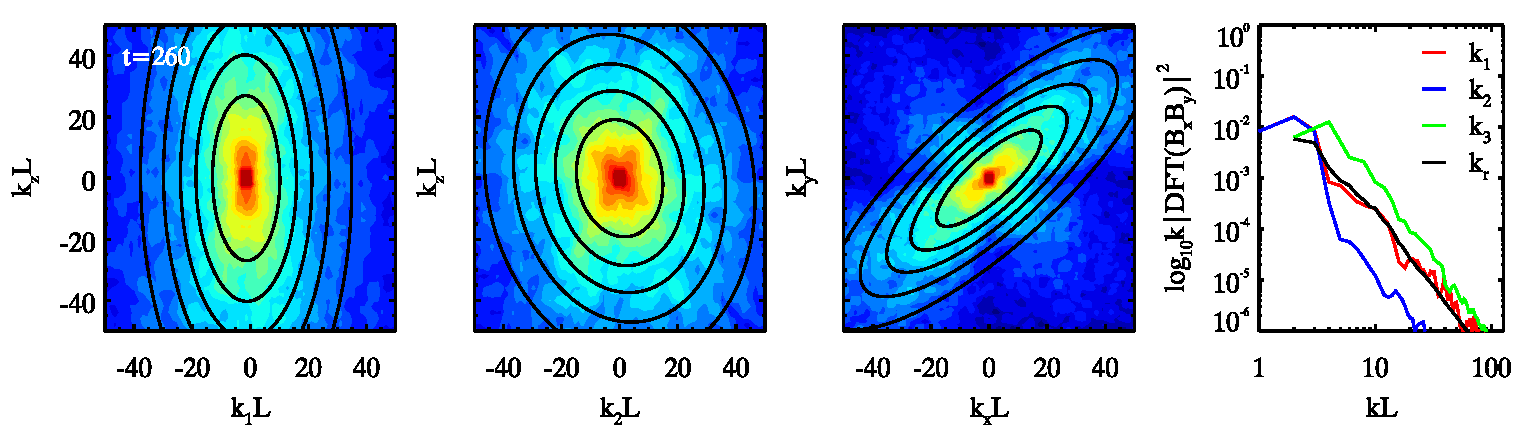}
\caption{
Analysis of the power spectral density of the Maxwell stress in the 
early stages of the nonlinear evolution of the MRI and the fully developed 
turbulent state. The upper and lower row correspond to $t=10~\Omega^{-1}$  
and $t=260~\Omega^{-1}$, respectively.
The first three panels in each row, from left to right, show slices of the 3D discrete 
Fourier Transform (DFT) of the Maxwell stress in the $(k_1,k_z)$,$(k_2,k_z)$ 
and $(k_x,k_y)$ planes, respectively. 
The black contours in each panel represent 2D gaussian elliptical fits to the 
smoothed distributions. The right-most column shows 1D cuts of the same data, compared with the spherical average as a function of the radial wave vector. 
The power spectra of the Maxwell stress exhibits a high degree of anisotropy, which depends 
on time.
}
\label{cutsof3dfftstress}
\end{center}
\end{figure*}

\subsection{Time-Dependent, Anisotropic Power Spectral Distributions}
\label{spectral}

A number of works have studied the Fourier spectrum of  MRI-driven turbulence
(see, e.g., \citealt{Hawley:1995gd} for early studies and \citealt{2010A&A...514L...5F,
Lesur:2011jh, 2014MNRAS.441.1855N} for more recent work). In order to visualize the distribution 
of power as a function of scale, it is customary to plot either spherical 
shell-averaged spectra or 2D slices in the planes perpendicular to the directions 
given by $\hat{\bb{k}}_x, \hat{\bb{k}}_y$, and $\hat{\bb{k}}_z$. 
The second approach reveals that the turbulence in highly 
anisotropic, rendering the first approach suspect. In either
case, temporal averages over many orbital time scales
are usually invoked in order to obtain smooth spectral.
However, as we have shown in the previous section, the degree of 
anisotropy of the turbulence varies with time significantly.
 
In order to gain a deeper understanding of the temporal evolution
of the anisotropy of the turbulence in Fourier space we pursue
a different approach as follows. We demonstrate the procedure by 
computing the Fourier spectrum\footnote{Standard Fourier analysis requires a periodic
domain, and the shearing box domain that we employ is only strictly
periodic in the azimuthal and vertical directions. The radial
direction, which is only shearing-periodic, becomes strictly periodic
only at specific times \citep{Hawley:1995gd}.  In order to compute
Fourier transforms at arbitrary times, it is necessary to first
transform in the $x$-direction, and then shift the appropriate phase
before taking the transform in the $y$ and $z$-directions. This
procedure is explained in detail in, e.g.,
\citet{2009MNRAS.397...52H}.
} associated with the magnetic energy 
density in the early stages after the saturation of the linear instability 
and in the turbulent state. It is useful to visualize the results by selecting 
two-dimensional cuts in the three-dimensional Fourier space.
The anisotropic nature of the turbulence in the $(k_x,k_y)$ plane presents a
natural direction to define two perpendicular directions in this plane.
We define $\hat{\bb{k}}_{1}$ as the unit wave-vector corresponding 
to the direction of maximum anisotropy in the plane spanned by $(k_x,k_y)$.
The unit vectors $\hat{\bb{k}}_{3} = \hat{\bb{k}}_{z}$ and $\hat{\bb{k}}_2 
= \hat{\bb{k}}_{3} \times \hat{\bb{k}}_{1}$,
complete the orthonormal triad.  Note that the directions provided by 
$\hat{\bb{k}}_{1}$ and $\hat{\bb{k}}_{2}$ depend in general on time!

Figure~\ref{cutsof3dfft} shows the two dimensional cuts along the
three different planes perpendicular to 
$\hat{\bb{k}}_{1}$, $\hat{\bb{k}}_{2}$, and $\hat{\bb{k}}_{z}$
for two different times, corresponding to the early times after the 
saturation of the linear phase of the MRI ($t=10~\Omega^{-1}$, upper row) 
and the fully developed turbulent state   ($t=260~\Omega^{-1}$,  lower row).
In order to quantify the degree of anisotropy, at a
given time, we performed a two-dimensional Gaussian fit, which is over
plotted in each panel.  The Gaussian fit
is carried out by identifying the peak of the power spectral density
and then fitting a one-dimensional Gaussian along the direction of
maximum anisotropy and another one-dimensional Gaussian
in the direction perpendicular to it. The width of these two Gaussians
determine the semi-major axis of the two-dimensional ellipses that are
plotted over the data.

At time $t=10~\Omega^{-1}$, the parasite mode has reached the peak of its 
super-exponential growth, as attested by the presence of 
significant power in non-axisymmetric modes.
The distribution of power in the  $(k_2,k_z)$ plane is mostly 
isotropic, the two-dimensional Gaussian fit has an aspect ratio 
of 1.12. On the other hand, the Gaussian fits performed in the 
$(k_1,k_z)$ and $(k_x,k_y)$ planes show aspect ratios of 
$\sim 1.88, 3.12$ respectively. 
In the fully developed turbulent state, at $t=260~\Omega^{-1}$, the flow is highly 
anisotropic in the $(k_1,k_z)$ plane. The aspect ratios of the Gaussian fits 
in the $(k_1,k_z)$, $(k_2,k_z)$ and $(k_x,k_y)$ planes are respectively 2.63, 1.16 and 3.59.
The anisotropy of the spectrum has therefore increased significantly
from the late linear evolution of the MRI to the ensuing turbulent state.

The contrast between the energy spectral density obtained using 
spherical shell-averaged values versus one-dimensional profiles 
in each of the three planes described above is depicted in the rightmost 
panel in Figure~\ref{cutsof3dfft}.  At early times, i.e., soon after
the parasitic modes have disrupted the highly organized flow 
driven by the exponential growth of the MRI, the one-dimensional
spectra along the directions $\hat{\bb{k}}_{1}$ and $\hat{\bb{k}}_{2}$
is not exceedingly different, although it is clear that there is more power 
along $\hat{\bb{k}}_{1}$. However, at later times, in the fully developed turbulent 
regime, the turbulent power along $\hat{\bb{k}}_{1}$ is almost two orders of 
magnitude higher than the power in the direction perpendicular to it 
throughout most of the scales. 
In this case, the shell average is unable to provide a good 
representation of the anisotropic power spectrum of the turbulence. 
In both cases, averaging over spherical shells masks the spread
of almost two orders of magnitude in power along the directions
given by $\hat{\bb{k}}_{1}$ and $\hat{\bb{k}}_{2}$.

The distribution of power characterizing the Maxwell stress in Fourier 
space can be analyzed in a similar way that we have analyzed the
magnetic energy density. The results are shown in Figure~\ref{cutsof3dfftstress}.
As observed for the magnetic energy density, the Maxwell stress is also
highly anisotropic and the degree of anisotropy varies significantly with time.
Note that the fast decrease in power along the  $k_x$ and $k_y$ directions, 
i.e., the usual directions along which 1D PSD cuts are plotted, is even steeper
for the Maxwell stress than for the magnetic energy.
This implies that if we are interested in characterizing the degree of anisotropy 
of the stress is even more critical to analyze the spectra along 
$\hat{\bb{k}}_{1}$ and $\hat{\bb{k}}_{2}$, which provide a better characterization
of the three-dimensional distribution of power in Fourier space.

\subsection{Defining Robust Averages In Fourier Space}
\label{robust}

It is evident that the distribution of power in Fourier space is highly anisotropic
and time dependent. Therefore, in order to provide a robust characterization of the turbulent
spectrum we proceed as follows. At a fixed time, we obtain the three dimensional 
PSD in Fourier space and define the direction of maximum anisotropy in horizontal 
planes. This is accomplished by fitting a 2D Gaussian ellipse to the distribution in 
the $(k_x,k_y)$ plane and defining the direction of maximum anisotropy
as $\bb{k}_1$ and the two orthogonal directions $\bb{k}_2$ and $\bb{k}_3$.
The one-dimensional power spectra along each of these directions are readily obtained.
This procedure is repeated for different times and the resulting one-dimensional spectra
can be properly averaged along each independent direction. 

\begin{figure}[t]
\begin{center}
\includegraphics[width= \columnwidth]{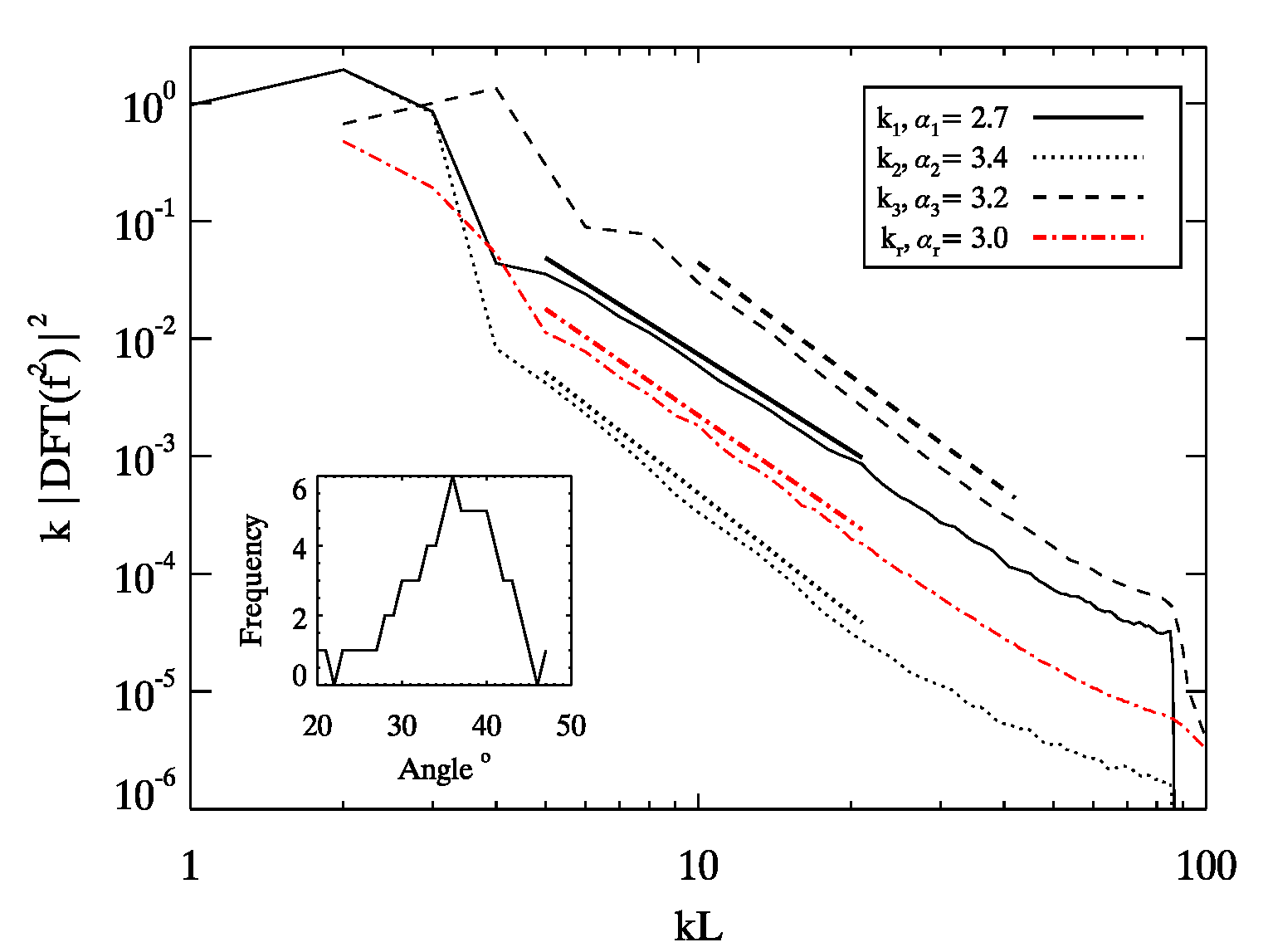}
\caption{
Time averaged plot of the distribution of energy in the directions given by 
$\hat{\bb{k}}_1$, $\hat{\bb{k}}_2$ and $\hat{\bb{k}}_3=\hat{\bb{k}}_z$ 
averaged over 260 $\Omega^{-1}$. The corresponding shell-averaged
spectrum is also shown for comparison. At high wave numbers the power
law behavior exhibited by $k|DFT(B^2)|^2$ can be described 
by $k^{-\alpha_i}$ with 
$\alpha_1 \sim 2.7, \alpha_2 \sim 3.4,  \alpha_z \sim 3$.
The index obtained for the shell-averaged spectrum is $\alpha_r = 3$.
The inset shows the distribution of values of the angle subtended by the 
direction of maximum anisotropy in the $(k_x, k_y)$ plane.
}
\label{angle}
\end{center}
\end{figure}

The average power spectra of the magnetic energy density obtained is this way is 
shown in Figure \ref{angle}.
Most of the power is concentrated in large scales and a power-law behavior
is observed beyond a characteristic wave number, which is different along
each direction.  In order to quantify the steepness of these profiles, we fitted  
power law  $k^{-\alpha_{i}}$ distributions to  $k|DFT(B^2)|^2$ along each 
independent direction $i=1,2,3$.
We obtained $\alpha_1 \sim 2.7, \alpha_2 \sim 3.4,  \alpha_z \sim 3$.
For comparison, we also plotted the power spectra that results from a standard
spherical shell average. This corresponds to $\alpha_r = 3$.
The orientation of the ellipsoid in Fourier space can be characterized by the angle
subtended between $\bb{k}_1$ and $\bb{k}_x$.  The average value of this angle 
is about $35^\circ$. The distribution of the values of this angle is shown in the inset.

The fact that the distribution of power in MRI-driven turbulence is anisotropic in Fourier space 
was already indicated by \citet{Hawley:1995gd}. Their Figure 4 shows that most of the power 
in the magnetic field resides in modes along $k_z$, in qualitatively agreement with our results.
\citet{Hawley:1995gd} also note a steepening at higher values of $k$ due to numerical diffusion, 
which we do not find in our simulations, possibly due to the incompressible pseudo-spectral 
method being less diffusive than the second order scheme (see, e.g., Figure 1 of \citealt{Brandenburg:2001us}.) 
Our results can also be compared with those of \citet{Lesur:2011jh}, who defined a spherical 
average and found a 3/2 slope for both the magnetic and kinetic energy spectra in the $1 < k < 10$ 
range, with a subsequent steeper falloff in the range $10<k<100$.
In those simulations, there does not seem to be clear sign of an inertial range.
The simulations in \citet{Lesur:2011jh} used a higher value of $\beta=q\Omega L/B^2=10^3$ 
than our $\beta \sim 10$, a slightly different aspect ratio of $4 \times 4 \times 1$, 
and used a broadband noise excitation. The fact that we have chosen the value of
$\beta$ such that a single wavelength of the fastest growing MRI mode fits in the
domain may have helped to set an effectively larger injection scale, possibly resulting
in the presence of a larger inertial range, as seen in Figure~\ref{angle}.

It is worth pointing out that most of the turbulent power, both in the magnetic energy
density and the Maxwell stress, resides in modes with wave vectors along $\bb{k}_z$, i.e., 
the direction perpendicular to the shear, with the contribution from non-axisymmetric modes
being subdominant. This could be due to the fact that we have considered a rather
strong vertical magnetic field. Whether this ordering of power in Fourier space is a 
generic feature of MRI-driven turbulence in more general settings, e.g., in the presence 
of weaker fields and/or more general magnetic field geometries, or stratification, will be 
addressed in subsequent work.

\section{Summary and Discussion}
\label{discussion}

In this paper, we have shed light into the  nature of MRI-driven
turbulence by studying in detail the temporal evolution of various indicators
that quantify the degree of anisotropy of the magnetized flow. We accomplished 
this task by starting with a well defined numerical experiment and analyzing
the outcome in ways that have not been previously  explored in the
context of MRI-driven turbulence.

We analyzed the saturation of the MRI via parasitic instabilities
and the subsequent transition to anisotropic turbulence by means of 3D
nonlinear MHD simulations carried out with the pseudo-spectral code
SNOOPY \citep{Lesur:2009ey}.  
We found that the exponential growth of the MRI is halted by the super-exponential
growth of a parasite instability, which corresponds to an
exponentially driven Kelvin-Helmholtz mode (see Section \ref{exact}).  
These findings are in general agreement with semi-analytical studies by
\citet{Goodman:1994dd, Pessah:2010ic} as well as the numerical studies by 
\citet{Latter:2009br,2010A&A...516A..51L,Obergaulinger:2009fv}.
Once the non-axisymmetric secondary instability reaches
an amplitude similar to the MRI mode, the highly organized,
anisotropic flow driven by the MRI becomes more isotropic and breaks
down into fully developed MHD turbulence (see Section \ref{latelinearMRI}).  

We showed that MRI-driven turbulence is
characterized by quasi-periodic fluctuations in the radial and
vertical velocity components throughout the duration of the
simulations (Figure \ref{fig:incompressiblemrigrowthrates400}).  
In particular, $v_x$ is tracked by $v_z$, with $v_z$
increasing until it reaches an amplitude equal to $v_x$, whereupon
they both decline together.  This behavior mimics the growth of $v_x$
followed by $v_z$ observed during the phase where parasitic
instabilities halt the growth of the MRI (Figure \ref{fig:incompressiblemrigrowthrates}).
This suggests that a turbulent version of the process that takes place
in the late stages of the linear regime of the instability could be at work.
This cyclic behavior is indicative of a strong temporal variation in the anisotropic
properties of the flow. We argued that more refined statistical tools
are needed in order to shed light into the anisotropic nature of MRI-driven
turbulence.  

We adopted a technique employed in hydrodynamic turbulence,
which is built on invariant analysis \citep{1977JFM....82..161L}, and 
used it to quantify and classify 
the anisotropy of MRI-driven turbulence (see Section \ref{lumley}). 
In particular, we extended the use of these tools to analyze the properties
of the Maxwell stress, which is the dominant source of angular
momentum transport in ideal MHD.  The invariant analysis shows a strong anisotropy,
with most of the power concentrated in the one-component regime, with
brief sporadic transitions towards two-component and three-component
turbulence.  We noted that the fact that the level of anisotropy fluctuates on
timescales comparable to $\sim\!\!10~\Omega^{-1}$  suggests that
averaging Fourier spectra over long time periods in order to obtain
smooth spectra, may influence the derived spectrum.

We also investigated the impact associated with using shell averages
to characterize the inherently anisotropic MRI-driven turbulence (see Section \ref{robust}).
This technique, which carries with it the implicit assumption of isotropy in 
Fourier space, has been widely adopted to characterize MRI-driven turbulence
\citep{Hawley:1995gd,Workman:2008cf,2010A&A...514L...5F}.
Our analysis consisted of dissecting the 3D Fourier spectrum, using 2D cuts
along planes selected according to the direction of maximum anisotropy. 
We found that the spectral state of turbulence is far from spherically symmetric
and that, in agreement with the invariant analysis, the degree of anisotropy
in Fourier space depends on time as well. 

This led us to propose, 
what we believe to be, a more robust way to average Fourier spectra,
by consistently stacking 1D spectra along the directions of maximum 
anisotropy at any given time. The result of this process is shown in
Figure~\ref{angle}, which also shows that the direction of maximum 
anisotropy in the plane $({k_x,k_y})$ oscillates around $35^\circ$,
with respect to $k_x$. The distribution of the angle subtended by 
the direction of maximum anisotropy, shown in the inset, reflects
the time-dependent nature of the anisotropy in Fourier space.

Using this procedure, we provided statistical significance to a fact that
was already evident in the spectra taken at two different times
(see, e.g., Figure~\ref{cutsof3dfft}). In the turbulent state, the power
along the three independent directions which show maximum anisotropy 
differs by several orders of magnitude over most 
scales, except the largest ones.  This indicates that spherically averaging
technique to seek scaling relationships for MRI-driven turbulence may
distort the underlying, inherently anisotropic turbulent spectrum. In particular, 
the identification of an inertial range for differentially rotating MHD turbulence, 
may be hampered by the use of spherical averages to seek for isotropic scaling laws, 
which are at odds with the strong anisotropy that leads naturally to efficient 
angular momentum transport.

\acknowledgments
The authors thank G. Lesur, C. McNally, O. Gressel, and J. Oishi for helpful discussions.
The results presented in this paper were partly achieved using the 
PRACE Research Infrastructure resource JUQUEEN based in Germany at 
the J\"ulich Supercomputing Centre. The authors also acknowledge DCSC/KU 
and ICHEC for computational facilities and support.
The research leading to these results has received funding from the 
European Research Council under the European Union's Seventh Framework 
Programme (FP/2007-2013) under ERC grant  agreement 306614 (MEP). 
The authors also acknowledge support from the Young Investigator 
Programme of the Villum Foundation.

\bibliography{MyBibFile}

\end{document}